%% file: draft.tex
\documentclass[longauth]{aa} 
\usepackage{multicol}
\usepackage{longtable}
\usepackage{url}
\usepackage{graphicx}
\usepackage{color}
\usepackage{txfonts} 
\usepackage[]{hyperref}

\hypersetup{breaklinks=true,colorlinks=true,linkcolor=blue,citecolor=blue,urlcolor=blue}

\usepackage{natbib,twoopt}
\bibpunct{(}{)}{;}{a}{}{,}             

\makeatletter
  \newcommandtwoopt{\citeads}[3][][]{\href{http://adsabs.harvard.edu/abs/#3}%
    {\def\hyper@linkstart##1##2{}%
     \let\hyper@linkend\@empty\citealp[#1][#2]{#3}}}
  \newcommandtwoopt{\citepads}[3][][]{\href{http://adsabs.harvard.edu/abs/#3}%
    {\def\hyper@linkstart##1##2{}%
     \let\hyper@linkend\@empty\citep[#1][#2]{#3}}}
  \newcommandtwoopt{\citetads}[3][][]{\href{http://adsabs.harvard.edu/abs/#3}%
    {\def\hyper@linkstart##1##2{}%
     \let\hyper@linkend\@empty\citet[#1][#2]{#3}}}
  \newcommandtwoopt{\citeyearads}[3][][]%
    {\href{http://adsabs.harvard.edu/abs/#3}
    {\def\hyper@linkstart##1##2{}%
     \let\hyper@linkend\@empty\citeyear[#1][#2]{#3}}}
\makeatother

\usepackage{longtable}

\graphicspath{{figures/}}

\usepackage{color}
\definecolor{red}{rgb}{0.7,0.1,0.1}
\definecolor{black}{rgb}{0.0,0.0,0.0}

\begin{document} 
\newcommand{\Ha}{$\rm{H}\alpha$}
\newcommand{\Hb}{$\rm{H}\beta$}
\newcommand{\logU}{$\log(\bar U)$}
\newcommand{\Kdist}{$K_{dist}$}
\newcommand{\EWa}{$EW_\alpha$}
\newcommand{\fOB}{f(OB)}
\newcommand{\QHHe}{Q$_{0/1}$}
\newcommand{\HII}{\ion{H}{ii}}
\newcommand{\Com}[1]{{\color{red}*** #1}}
\newcommand{\temp}{T$_{\rm e}$}


\newcommand{\kms}{km\,s$^{-1}$} 
\newcommand{\hi}{\ion{H}{i}}
\newcommand{\hh}{\ion{H}{ii}}
\newcommand{\nii}{[\ion{N}{ii}]}
\newcommand{\sii}{[\ion{S}{ii}]}
\newcommand{\siii}{[\ion{S}{iii}]}
\newcommand{\oi}{[\ion{O}{i}]}
\newcommand{\oii}{[\ion{O}{ii}]}
\newcommand{\oiii}{[\ion{O}{iii}]}
\newcommand{\ha}{H$\alpha$} 
\newcommand{\hb}{H$\beta$} 
\newcommand{\flux}{erg\,s$^{-1}$\,cm$^{-2}$}


\titlerunning{The CALIFA survey IV. Third public data release}
\authorrunning{Sanchez et al.}
\title{CALIFA, the Calar Alto Legacy Integral Field Area survey}
\subtitle{IV. Third Public data release\thanks{Based on observations collected 
at the Centro Astron\'omico Hispano Alem\'an (CAHA) at Calar Alto, operated jointly 
by the Max-Planck-Institut f\"ur Astronomie (MPIA) and the Instituto de Astrof\'isica de
Andaluc\'ia (CSIC)}}

\author{S.~F.~S\'anchez\inst{\ref{unam}}
   \and R.~Garc\'ia-Benito\inst{\ref{iaa}}
   \and S.~Zibetti\inst{\ref{floren}}
   \and C.~J.~Walcher\inst{\ref{aip}}
   \and B.~Husemann\inst{\ref{eso}}
   \and M.~A.~Mendoza\inst{\ref{iaa}}   
   \and L.~Galbany\inst{\ref{mil},\ref{chile}}
\and J.~Falc\'on-Barroso\inst{\ref{iac},\ref{lagu}}
   \and D.~Mast\inst{\ref{cordoba},\ref{cordoba1}}
   \and J.~Aceituno\inst{\ref{caha}}
   \and J.~A.~L.~Aguerri\inst{\ref{iac}}
   \and J. Alves\inst{\ref{vienna}}
   \and A.~L.~Amorim\inst{\ref{flori}}
   \and Y.~Ascasibar\inst{\ref{uam}}
   \and D.~Barrado-Navascues\inst{\ref{caha},\ref{bio}}
   \and J.~Barrera-Ballesteros\inst{\ref{jhop}}
   \and S.~Bekerait\`e\inst{\ref{aip}}
   \and J.~Bland-Hawthorn\inst{\ref{sydney}}
   \and M.~Cano~D\'\i az\inst{\ref{unam}}
   \and R.~Cid Fernandes\inst{\ref{flori}}
   \and O.~Cavichia\inst{\ref{itajuba}}
   \and C.~Cortijo\inst{\ref{iaa}}
   \and H.~Dannerbauer\inst{\ref{iac}}
   \and M.~Demleitner\inst{\ref{zent}}
   \and A.~D\'iaz\inst{\ref{uam}}
   \and R.~J.~Dettmar\inst{\ref{bochum}}
   \and A.~de Lorenzo-C\'aceres\inst{\ref{stAndrews},\ref{ugr}}
   \and A.~del Olmo\inst{\ref{iaa}}
   \and A.~Galazzi\inst{\ref{floren}}
   \and B.~Garc\'\i a-Lorenzo\inst{\ref{iac}}
   \and A.~Gil~de~Paz\inst{\ref{ucm}}
   \and R.~Gonz\'alez Delgado\inst{\ref{iaa}}
   \and L.~Holmes\inst{\ref{kings}}
   \and J.~Igl\'esias-P\'aramo\inst{\ref{iaa}}
   \and C.~Kehrig\inst{\ref{iaa}}
   \and A. Kelz\inst{\ref{aip}}
   \and R.~C.~Kennicutt\inst{\ref{IoA}}
   \and B.~Kleemann\inst{\ref{bochum}}
   \and E.~A.~D.~Lacerda\inst{\ref{flori}}
   \and R.~L\'opez Fern\'andez\inst{\ref{iaa}}
   \and A.~R.~L\'opez S\'anchez\inst{\ref{sydney}}
   \and M.~Lyubenova\inst{\ref{kapteyn}}
   \and R.~Marino\inst{\ref{ETH}}
   \and I.~M\'arquez\inst{\ref{iaa}}
   \and J.~Mendez-Abreu\inst{\ref{stAndrews}}
   \and M.~Moll\'a\inst{\ref{ciemat}}
   \and A.~Monreal-Ibero\inst{\ref{meudon}}
   \and R.~Ortega Minakata\inst{\ref{rio}}
   \and J.~P.~Torres-Papaqui\inst{\ref{guana}}
   \and E.~P\'erez\inst{\ref{iaa}}
   \and F.~F.~Rosales-Ortega\inst{\ref{inaoe}}
   \and M.~M.~Roth\inst{\ref{aip}}
   \and P.~S\'anchez-Bl\'azquez\inst{\ref{uam},\ref{chile_pat}}
   \and U.~Schilling\inst{\ref{bochum}}
   \and K.~Spekkens\inst{\ref{kings}}
   \and N.~Vale~Asari\inst{\ref{flori}}
   \and R.~C.~E.~van den Bosch\inst{\ref{mpia}}
   \and G.~van de Ven\inst{\ref{mpia}}
   \and J.~M.~Vilchez\inst{\ref{iaa}}
   \and V.~Wild\inst{\ref{stAndrews}}
   \and L.~Wisotzki\inst{\ref{aip}}
   \and A.~Y{\i}ld{\i}r{\i}m\inst{\ref{mpia}}
   \and B.~Ziegler\inst{\ref{vienna}}
}

\institute{Instituto de Astronom\'ia, Universidad Nacional Auton\'oma de M\'exico, A.P. 70-264, 04510, M\'exico, D.F.\label{unam}\email{sfsanchez@astro.unam.mx}
  \and Instituto de Astrof\'isica de Andaluc\'ia (IAA/CSIC), Glorieta de la Astronom\'{\i}a s/n Aptdo. 3004, E-18080 Granada, Spain,\label{iaa} 
  \and INAF-Osservatorio Astrofisico di Arcetri - Largo Enrico Fermi, 5 - I-50125 Firenze, Italy\label{floren} 
  \and Leibniz-Institut f\"ur Astrophysik Potsdam (AIP), An der Sternwarte 16, D-14482 Potsdam, Germany\label{aip}
  \and European Southern Observatory, Karl-Schwarzschild-Str. 2, D-85748 Garching b. M{\"u}nchen, Germany\label{eso}
  \and Millennium Institute of Astrophysics, Universidad de Chile, Santiago, Chile\label{mil}
  \and Departamento de Astronom\'ia, Universidad de Chile, Casilla 36-D, Santiago, Chile\label{chile}
  \and Instituto de Astrof\'isica de Canarias, V\'ia L\'actea s/n, La Laguna, Tenerife, Spain\label{iac}
  \and University of Vienna, Department of Astrophysics, T\"urkenschanzstr. 17, 1180 Vienna, Austria\label{vienna}
  \and Departamento de Astrof\'isica, Universidad de La Laguna, E-38205 La Laguna, Tenerife, Spain\label{lagu}
  \and Observatorio Astron\'omico, Laprida 854, X5000BGR, C\'ordoba, Argentina\label{cordoba}
  \and Consejo de Investigaciones Cient\'{i}ficas y T\'ecnicas de la Rep\'ublica Argentina, Avda. Rivadavia 1917, C1033AAJ, CABA, Argentina\label{cordoba1}
  \and Centro Astron\'omico Hispano Alem\'an de Calar Alto (CSIC-MPG), C/ Jes\'us Durb\'an Rem\'on 2-2, E-4004 Almer\'ia, Spain\label{caha}
  \and Departamento de F\'{\i}sica, Universidade Federal de Santa Catarina, P.O. Box 476, 88040-900, Florian\'opolis, SC, Brazil\label{flori} 
  \and Centro de Astrobiología (CSIC-INTA), Depto. Astrof\'isica, ESAC Campus, 28691 Villanueva de la Ca\~nada, Madrid, Spain \label{bio}
  \and Department of Physics \& Astronomy, Johns Hopkins University, Bloomberg Center, 3400 N. Charles St., Baltimore, MD 21218, USA\label{jhop}
  \and Sydney Institute for Astronomy, School of Physics, University of Sydney, NSW 2006, Australia\label{sydney}
  \and Instituto de F\'isica e Qu\'imica, Universidade Federal de Itajub\'a, Av. BPS, 1303, 37500-903, Itajub\'a-MG, Brazil\label{itajuba}
  \and Universit\"at Heidelberg, Zentrum f\"ur Astronomie, Astronomisches Rechen-Institut, M\"onchhofstra{\ss}e 12-14, D-69120 Heidelberg, Germany\label{zent}
  \and Departamento de F\'isica Te\'orica, Facultad de Ciencias, Universidad Aut\'onoma de Madrid, E-28049 Madrid, Spain\label{uam} 
  \and Astronomisches Institut, Ruhr-Universit{\"a}t Bochum, Universit{\"a}tsstr. 150, D-44801 Bochum, Germany\label{bochum}
  \and School of Physics and Astronomy, University of St Andrews, SUPA, North Haugh, KY16 9SS, St Andrews, UK \label{stAndrews}
  \and Departamento de F\'{\i}sica Te\'orica y del Cosmos, University of Granada, Facultad de Ciencias (Edificio Mecenas), E-18071 Granada, Spain\label{ugr}
 \and Departamento de Astrof{\'i}sica y CC. de la Atm{\'o}sfera, Universidad Complutense de Madrid, E-28040, Madrid, Spain\label{ucm}
  \and Department of Physics, Royal Military College of Canada, PO box 17000, Station Forces, Kingston, Ontario, Canada, K7K 7B4\label{kings}
  \and Institute of Astronomy, University of Cambridge, Madingley Road, Cambridge CB3 0HA, UK\label{IoA}
  \and Kapteyn Astronomical Institute, University of Groningen, Postbus 800, 9700 AV Groningen, The Netherlands\label{kapteyn}
  \and Institut f\"ur Astronomie (ETH), Wolfgang-Pauli-Str. 27, 8093 Zürich, Switzerland\label{ETH}
  \and CIEMAT, Avda. Complutense 40, 28040 Madrid, Spain\label{ciemat}
  \and GEPI, Observatoire de Paris, CNRS, Université Paris Diderot, Place Jules Janssen, 92190 Meudon, France\label{meudon}
  \and Observatório do Valongo, Universidade Federal do Rio de Janeiro, Ladeira do Pedro Antônio 43, Saúde, Rio de Janeiro, RJ 20080-090, Brazil\label{rio}
  \and Departamento de Astronomía, Universidad de Guanajuato, Apartado Postal 144, 36000, Guanajuato, Guanajuato, Mexico\label{guana}
  \and Instituto Nacional de Astrof{\'i}sica, {\'O}ptica y Electr{\'o}nica, Luis E. Erro 1, 72840 Tonantzintla, Puebla, Mexico\label{inaoe}
  \and Instituto de Astrof\'{\i}ısica, Pontificia Universidad Cat\'olica de Chile, Av. Vicu\~na Mackenna 4860, 782-0436 Macul, Santiago, Chile\label{chile_pat}
  \and Max-Planck-Institut f\"ur Astronomie, K\"onigstuhl 17, D-69117 Heidelberg, Germany\label{mpia}}

\date{}

 
  \abstract
    {This paper describes the Third Public Data Release (DR3) of the Calar Alto Legacy Integral Field Area 
   (CALIFA) survey. Science-grade quality data for 667 galaxies are made public, including the 200 galaxies 
   of the Second Public Data Release (DR2). Data were obtained with the integral-field spectrograph PMAS/PPak
   mounted on the 3.5m telescope at the Calar Alto Observatory. Three different spectral setups are available, 
   i) a low-resolution V500 setup covering the wavelength range 3745–7500 \AA\ (4240-7140 \AA\ unvignetted) 
   with a spectral resolution 
   of 6.0 \AA\ (FWHM), for 646 galaxies, ii) a medium-resolution V1200 setup covering the wavelength range 
   3650–4840 \AA\ (3650-4620 \AA\ unvignetted) with a spectral resolution of 2.3 \AA\ (FWHM), for 484 
   galaxies, and iii) the combination of 
   the cubes from both setups (called COMBO), with a spectral resolution of 6.0 \AA\ and a wavelength 
   range between 3700-7500 \AA\ (3700-7140 \AA\ unvignetted), for 446 galaxies. 
   The Main Sample, selected and observed according to the CALIFA survey strategy covers a redshift range between 
   0.005 and 0.03, spans the color-magnitude diagram and probes a wide range of stellar mass, ionization 
   conditions, and morphological types. 
   The Extension Sample covers several types of galaxies that are rare in the overall 
   galaxy population and therefore not numerous or absent in the CALIFA Main Sample. 
   All the cubes in the data release were processed using the latest pipeline, which includes improved 
   versions of the calibration frames and an even further improved image reconstruction quality.
   In total, the third data release contains 1576 datacubes, including $\sim$1.5 million independent spectra. 
   It is available at \url{http://califa.caha.es/DR3}.}

   \maketitle

\section{Introduction}


The advent of large imaging surveys, complemented in some cases by single-fiber spectroscopy 
\citep[e.g., Sloan Digital Sky Survey, SDSS, Galaxy and Mass 
Assembly survey, GAMA][]{york2000,gamma}, has opened important new avenues for our understanding of 
galaxy evolution. However, one of the most significant limitations of these surveys is that they 
do not provide resolved spectroscopic information.
Galaxies have long been known to be spatially extended objects, with observed properties that vary 
across their optical extents \citep[e.g.~][]{hubble26,hubble36}. Many of these properties vary 
coherently as a function of position relative to the galaxy centre, and radial gradients have 
been studied for decades \citep[e.g.~][]{pagel81,rene89}. Characterizing galaxies by assigning 
global values therefore over-simplifies their true complexity, and determining the spatially 
resolved properties of galaxies is fundamental to understanding the evolutionary processes that 
have shaped them. At the same time, evidence for a diversity in galaxy evolutionary histories as 
a function of mass and environment implies that statistically significant samples over large 
fractions of the sky are needed to understand the underlying physical mechanisms at work.

Thus the logical next step for pushing beyond multi-band imaging surveys (that provide detailed 
spatial information and limited spectral information) or single-aperture spectroscopic surveys 
(that sample only limited galaxy regions) is an Integral Field Spectroscopy (IFS) survey over a 
representative and statistically significant sample of galaxies. With this aim we undertook the 
Calar Alto Legacy Integral Field Area (CALIFA) survey in 2010, \citep[][]{Sanchez:2012a} to obtain 
spatially resolved spectra for $ \sim$600 galaxies in the local universe.

{

CALIFA was the first survey using imaging spectroscopy that was meant from the outset to provide 
a public dataset of a sample of galaxies representative of the Local Universe, i.e. CALIFA was 
from the outset foreseen to be a legacy survey. CALIFA thus aimed at extending the pioneering SAURON (Spectrographic Areal Unit for Research on Optical Nebulae) 
and Atlas3D surveys \citep{cappellari10ar} to all galaxy types and larger wavelength coverage. 
The next generation of IFS surveys are already going on -- MaNGA \citep[Mapping Nearby Galaxies at Apache Point Observatory survey, ][]{manga} and SAMI \citep[Sydney Australian-Astronomical-Observatory Multi-object Integral-Field-Spectrograph survey,][]{sami}. 
It is beyond the scope of this article to make a detail comparison between those surveys and CALIFA, 
a topic that has already been addressed in previous articles \citep[e.g.][]{2015IAUS..309...85S}. 
Briefly, we note that MaNGA and SAMI will supplant all previous surveys in terms of number of objects. 
They have adopted a multiplexing scheme that allows to observe several objects simultaneously. This 
increases the efficiency of collecting data in terms of number of objects which was their main goal 
(10,000 objects, for MaNGA, and 3,600 objects, for SAMI). CALIFA and Atlas3D use a single-IFU mode, 
which limits the number of objects to be observed simultaneously. On the other hand the latter surveys 
observe a larger number of spectra per object, and offer better physical spatial sampling. One obvious 
manifestation of this is the very similar total number of spectra obtained by CALIFA, MaNGA and SAMI, 
as compared to the very different total number of objects to be observed. 

The sample selections are also quite different. The MaNGA and SAMI samples cover over a wider range 
of redshifts, from $z\sim$0.001 to $z\sim$0.16. Because the Full Width at Half Maximum (FWHM) of the 
Point Spread Function (PSF) is very similar between the three surveys this implies a wide range of 
physical resolutions. Galaxies are thus sampled over a wider range of masses, but at different 
cosmological distances. On the contrary, the redshift range of the CALIFA and Atlas3D samples are 
rather small \citep{cappellari10ar,walcher14}, and therefore they present a better and more uniform 
physical sampling, making them optimal to study spatially resolved structures in galaxies (at $\sim$1 kpc 
resolution). 

Another important difference is the coverage of the different IFU surveys in terms of the optical 
extension of the galaxies. CALIFA observations cover most of the optical extension (beyond 
$\sim$2.5 $r_e$) by construction, while Atlas3D reaches between 1-1.5$r_e$ on average. MaNGA comprises 
two main samples, where the goal is to reach either 1.5$r_e$ ($\sim$70\% of the targets) or 2.5$r_e$ 
($\sim$20\%), and it hardly samples the outer regions for most of the galaxies (Ibarra-Mendel et al., 
submitted). Finally, SAMI, with the smallest FoV of all IFU surveys (16$\arcsec$/D) covers around 
$\sim$1$r_e$ of the galaxies \citep{Bryant:2014}.

In summary the data provided by the CALIFA DR3 presented here occupy a niche which ensures high spatial 
resolution and good spatial coverage simultaneously, however, at the price of a smaller sample in 
comparison with currently ongoing surveys like MaNGA and SAMI. } The CALIFA collaboration has addressed 
many different science cases using the information provided by these 
data, all of them focused on understanding the main properties of galaxies in the local universe and the 
evolutionary processes that have shaped them: i) New techniques have been developed to
understand the spatially resolved star formation histories (SFH) of
galaxies \citep{CidFernandes:2013,CidFernandes:2014,lopez2016}. Clear evidence that mass-assembly in 
typical galaxies progresses inside-out \citep{perez13}. The SFH and chemical
enrichment of bulges and early-type galaxies are fundamentally related
to their total stellar mass, while for disk galaxies they are more closely related
to the local stellar mass density \citep{rosa14,rosa14b,rosa15a}. Negative age
gradients indicate that quenching is progressing outwards in 
massive galaxies \citep{rosa15a}, and age and metallicity gradients
suggest that galaxy bars have not significantly altered the SFHs of
spirals \citep{patri14}. Finally, we explored spatially resolved
stellar populations and star formation across the Hubble sequence
\citep{rosa15a,rosa16}, and how mergers influence the assembly of blue
ellipticals \citep{haines15}. ii) We studied the origin of the low
intensity, LINER-like, ionized gas in galaxies. These regions are
clearly not related to recent star-formation activity, or to AGN activity
\citep{sign13}. They are most probably related to post-AGB ionization
in many cases \citep{kehrig12,papa13,gomes15}. iii) We explored aperture and resolution effects affecting 
larger single-fiber (like SDSS) and IFS surveys \citep[like MaNGA and 
SAMI][]{manga,sami}. 
We explored the effects of signal dilution in IFS data obtained for higher redshift galaxies in different gas 
and stellar population properties \citep{Mast:2014}, and proposed a new empirical aperture correction for
SDSS data \citep{IglesiasParamo:2013,jiglesias16}. We also compared average stellar and ionized gas 
properties with spatially resolved ones \citep[e.g][]{rosa14,rosa15a}; iv) We studied the kinematics of
the ionized gas \citep{GarciaLorenzo:2014}, the effects of bars on the
kinematics of galaxies \citep{BarreraBallesteros:2014a,holmes15}, the effects of the
interaction stage on the kinematic signatures \citep{jkbb15}, 
and measured the bar pattern speeds in late-type galaxies
\citep{Aguerri:2015}. v) We extended measurements of the angular
momentum of galaxies to previously unexplored ranges of morphology and
ellipticity \citep[e.g.~][]{barroso15} and proposed a new dynamical
classification scheme for galaxies \cite{kali15}.
The stellar dynamics together with detail analysis of the stellar populations revealed a tight 
relation between the Initial Mass Function (IMF) and the local metallcity \citep{navarro16} and 
allowed us to dynamically constrain the shape of the IMF in early-type galaxies (Lyubenova et al. 
submitted). vi) We explored in
detail the impact of galaxy interactions on the enhancement of
star-formation rates and the ignition of galactic outflows \citep{wild14,jkbb15b}. vii) We studied the nature 
of the progenitors of SNe \citep{Galbany:2014}; viii) we explored 
star formation indicators for extended objects and the suitability of
H$\alpha$ as a star-formation rate (SFR) tracer \citep{catalan15}, as well as the spatially resolved 
SFR and SFR density
across the Hubble sequence \citep{rosa16}. 
ix) We studied oxygen abundance gradients in the gas, developing new 
calibrators \citep{Marino:2013}, finding a characteristic shape \citep{Sanchez:2014,Laura16}, and a weak 
dependence of the profile truncations on the gradient properties \citep{marino16} and 
the stellar populations \citep{ruiz16}. x) We explored the mass -- metallicity relation for both the stellar 
populations \citep{rosa14b} and the gas oxygen abundance \citep{Sanchez:2013}. We could not confirm a 
secondary relation between the SFR and the Metallicity \citep{sanchez15Y}. xi) Finally, we found that many of 
the global scaling relations such as the star formation main sequence or the 
mass--metallicity relation are mirrored by local relations that hold on a scale of $\sim1$kpc 
\citep[e.g.][]{Sanchez:2013,cano16}.

CALIFA was designed as a legacy survey, and therefore we have distributed the data in successive Data 
Releases (DR) as the number of observed objects has increased and the processing pipeline has improved 
\citep[DR1 and DR2,][respectively]{Husemann:2013,rgb15}. These publicly accessible data have already allowed the
exploration of several different scientific avenues not addressed by
the collaboration \citep[e.g.,][]{Holwerda:2013,DeGeyter:2014,MartinezGarcia:2014,Davies:2014,roche15, ho15}.
In this article we present the third and final Data Release (DR3) comprising all observations with good quality. 
We distribute 1576 datacubes corresponding to 667 galaxies, 646 of them observed with the V500 setup, 
484 observed with the V1200 setup and 446 combined (COMBO) cubes.

The properties of the galaxies in the DR3 sample are described in
Section \ref{sect:DR3_sample}.  We describe the observing strategy and
setup (Section \ref{sect:observations}), processing
(Section \ref{sect:data_processing}), structure
(Section \ref{sect:data_format}), and data (Section \ref{sect:QC}), which
comprise essential information for any scientific analysis of the
distributed CALIFA data.  Several interfaces to access the CALIFA DR3
data are explained in Section \ref{sect:DR2_access}.


\section{The CALIFA DR3 sample}
\label{sect:DR3_sample}

There are two fundamentally different samples of galaxies in the CALIFA DR3: (1) Galaxies that were targeted because they were part of the CALIFA 
mother sample that is fully described and characterized in \citet{Walcher:2014}. This sample is called the CALIFA Main Sample. 
(2) The CALIFA Extension Sample, which is a heterogeneous set of galaxies observed for various reasons 
as part of different ancillary science projects within the CALIFA collaboration. 
The DR3 release is the combination of the Main Sample and the Extension Sample.

\subsection{Main sample}

The Main Sample contains all galaxies for which cubes are released in the CALIFA DR3 and that have been 
drawn from the CALIFA mother sample. 

\subsubsection{Mother sample - a recap}

The CALIFA mother sample consists of 937 galaxies drawn from SDSS DR7. In earlier papers 
\citet[in particular][]{Walcher:2014} we quoted the number 939, because we included the galaxies NGC5947 and NGC4676B. 
However, these galaxies did not formally satisfy the selection criteria of the mother sample but 
were added by hand. We therefore now shifted them to the Extension Sample, where they have a natural 
place. 

The main criteria for the target selection of the mother sample are fully described in \citet{Walcher:2014}. 
Briefly for completeness, they are:
\begin{itemize}
\item angular isophotal diameter $45\arcsec < $isoA$_{r} < 79.2\arcsec$; 
\item redshift range $0.005<z<0.03$; 
\item Galactic latitude $|b| > 20\degr$; 
\item flux limit petroMag$_{r} < 20$; 
\item declination $\delta > 7\degr$. 
\end{itemize}

The lower redshift limit was imposed so that the mother sample would not be dominated 
by dwarf galaxies. CALIFA thus has a natural lower completeness limit in mass. The upper redshift limit was 
imposed to keep relevant spectral features observable with a fixed instrumental 
setup. This limits the total volume probed by the CALIFA sample to roughly $10^6$ Mpc$^3$. Because 
massive galaxies are very rare, this volume effectively sets the upper mass boundary of the CALIFA sample 
(and not the diameter selection). 
The 95\% completeness limits of the mother sample are studied in detail in \citet{Walcher:2014} and are as 
follows: -19 $>$ M$_{r}$ $>$ -23.1 in luminosity and 10$^{9.7}$ and 10$^{11.4}$ M$_{\odot}$ in 
stellar mass \citep[with a Chabrier Initial Mass Function,][]{chab03}. 

In \citet{Walcher:2014} we showed that the mother sample has well understood properties. 
In particular, the diameter selection 
can be translated into V$_{\mathrm{max}}$ values according 
to the formalism of \citet{schmidt68}. This allows us to construct the luminosity function from 
the mother sample and to show that it agrees with the standard literature determination of the 
luminosity function within the limits imposed by the sample size. Indeed, while the selection effects are 
understood and can be corrected within our completeness limits, the 
finite sample size of CALIFA still implies that some galaxy classes have less representatives within the sample. 
Specifically for the CALIFA Main Sample, the statistics are best for galaxies with stellar masses 
around $10^{10.8}$ M$_{\odot}$.

\subsubsection{Definition of the Main Sample}

Galaxies were selected for observation from the mother sample randomly, i.e.~based only on visibility. 
We can therefore assume that the Main Sample is a random subset of the mother sample. 
Below we will proceed to verify this assumption. We base our verification on the same galaxy physical 
properties studied in \citet{Walcher:2014}. For DR3 we have re-compiled the catalogues of physical
properties for 
two reasons: (1) We introduced a few bug fixes in column names or with single numbers in the catalogues. 
(2) We computed new properties based on SDSS Petrosian magnitudes to allow for comparison with the 
Extension Sample. All tabulated properties are available for all galaxies of the mother sample, 
i.e.~by definition for all Main Sample galaxies in the release. 

The Main Sample as used below contains all galaxies with either a V500 and/or a V1200 cube released in this 
data release and no quality control flags that mark them as unusable (see Section \ref{sect:QC_select}). 
The number of galaxies in the Main Sample is 542. 

\subsection{Extension Sample}

The Extension Sample consists of an inhomogeneous collection of galaxies observed in the CALIFA setup, but not 
following the same sample selection criteria of the mother sample. This means in particular, that V$_{\mathrm{max}}$ 
values cannot be computed for the Extension Sample. On the other hand, Extension Sample galaxies have mostly been 
selected to cover galaxies that are intrinsically rare and thus not found in the CALIFA Main Sample. They 
thus provide useful benchmarks for such rare galaxies. 

The CALIFA setup is used in the observations of 
the Extension Sample galaxies, i.e.~the same gratings, grating angles, 
exposure times and observing strategy. However, many extension programs did not select galaxies from the SDSS DR7 
imaging survey. This imaging is needed to ensure good photometric calibration (Section \ref{sect:reducview}). 
Thus, for all extension programs, the additional selection criteria of being in the SDSS footprint and of 
fulfilling the Quality Control (QC) requirements (Section \ref{sect:QC}) are imposed. 

\subsubsection{Dwarf galaxies}

The extension program on dwarf galaxies is led by Garc\'ia-Benito. 
The galaxies in this program have CALIFA IDs between 1000 and 1999. 
The project aims to observe a statistically meaningful sample of dwarf galaxies in the local universe 
(9 < D < 40 Mpc). The sample was selected to be a magnitude-limited sample of local field galaxies observed 
with SDSS and covering a similar observed magnitude range as the CALIFA Main Sample.

The following selection criteria were used: i) The size of the galaxy (optical diameters) fill the PPak FoV, 
i.e.~29.7$"<$isoA$_r<79.2"$; ii) The ratio of the minor to major axis isoB$_r$/isoA$_r>0.6$, so that the galaxies 
are found to be almost face on; iii) M$_{r}$ $> -18$ iv) $z >$ 0.002. 

The input sample contained a total of 82 objects of which 34 were observed and included in the CALIFA DR3.

\begin{figure*}
\resizebox{0.49\hsize}{!}{\includegraphics{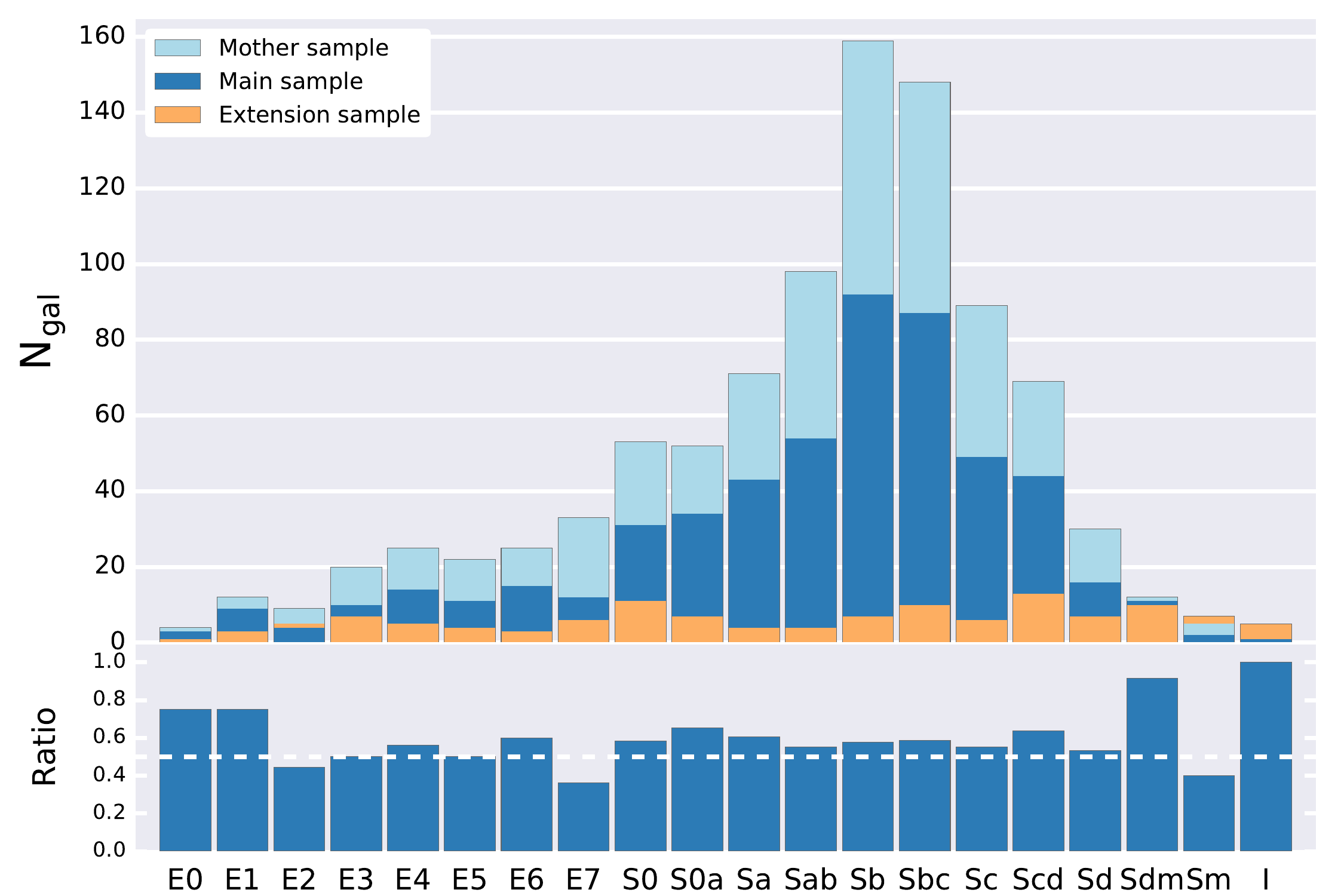}}
\resizebox{0.49\hsize}{!}{\includegraphics{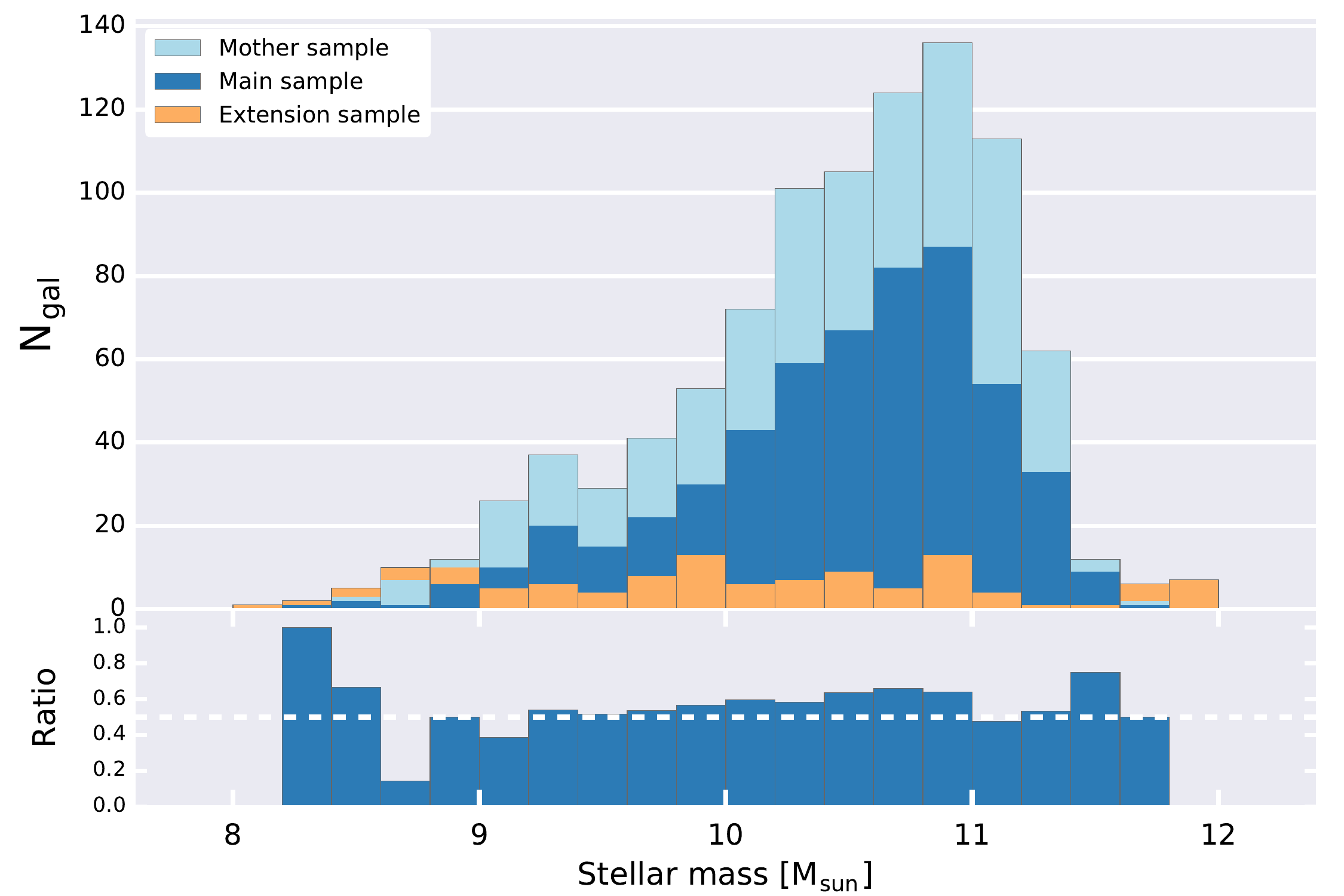}}
  \caption{\emph{Left panel:} Histograms of visual morphological classification in the DR3 samples. 
 \emph{Right panel:} Stellar mass histogram. 
 The lower portion of each panel shows the ratio between the Main Sample and the mother sample. }
  \label{fig:morph_mstar}
\end{figure*}

\subsubsection{Pairs and interacting galaxies}

The extension program on pairs and interacting galaxies is led by Barrera-Ballesteros. 
The galaxies in this program have CALIFA IDs between 2000 and 2999. 
The project aims to complete the IFU data for those pairs where only one companion galaxy was included 
in the CALIFA mother sample. The sample was selected to include companions of a CALIFA 
Main Sample galaxy with: i) a projected distance of 160 kpc; ii) a systemic velocity difference of less than 300 km/s; ii) an $r$-band magnitude difference of less than 2 mag. For details of the CALIFA pair selection 
see \citet{jkbb15b}. 

This sample also includes the second galaxy of the "mice" (NGC4676B) as ID number 2999. 
This galaxy was observed in the time allocated to the main survey since it seemed worthwhile to cover the full 
"mice" merger. These data have been published in \citep[][]{wild14}. 

In total there are 29 galaxies from this program in the CALIFA DR3. 

\subsubsection{Low- and high-mass early type galaxies}

The extension program on early-type galaxies (ETGs) is led by van de Ven, Lyubenova \& Meidt. 
The galaxies in this program have IDs between 3000 and 3999. 
This extension contains are three subprograms with the following scientific aims: a) studying the dark matter content of low-mass ETGs, 
b) constraining the initial mass function in high-mass ETGs, and c) testing fading scenarios for low-mass early-type spirals. 

The selection criteria for the low-mass ETGs were: a) $9.4<$ log(M*/M$_{\odot}$) $<10.4$; b) $35"<$isoA$_r<79.2"$; 
c) Declination $\delta>7^{\circ}$ and $75^{\circ} <$ RA $< 300^{\circ}$; and d) concentration $r_{90}$/$r_{50}>2.8$; visual inspection 
to remove non-ETGs. The selection criteria for the high-mass ETGs were: a) log(M*/M$_{\odot}$) $>11.4$; b) 
$35" < $isoA$_r$ $< 79.2"$; c) Declination $\delta> 7^{\circ}$ and $75^{\circ} <$ Right Ascension $RA<300^{\circ}$; d)
concentration $r_{90}$/$r_{50} > 2.8$; and e) visual inspection to remove non-ETGs. 
The selection criteria for low-mass early-type 
spirals were: a) $9.0 <$ log(M*/M$_{\odot}$) $< 10.0$; b) $30"<$isoA$_r<80"$; c) Declination $\delta>7^{\circ}$; 
d) inclination = acos(1-isoB$_r$/isoA$_r$) $< 80^{\circ}$; e) morphological types S0, Sa and Sb; 
and f) preference to those with literature HI observations. 

For all three subsamples the number of available galaxies was then reduced further by visibility at the scheduled time 
of observation. There are 36 galaxies from this program in the CALIFA DR3. 

\subsubsection{Pilot studies}

Those galaxies with IDs in the range 4000-4999 were extracted from CALIFA pilot studies and are fully 
described in \citet{MarmolQueralto:2011}. Most of the pilot study 
galaxies were observed with either the V300 or the V600 gratings and can therefore not be included in this 
homogeneous data release. The pilot studies targeted spiral galaxies with low inclinations 
to study the properties of the HII regions, as described in \citet{Sanchez:2012b}. 

Note that the galaxy with ID number 4034, NGC5947, is a galaxy from 
the pilot study sample that was observed as part of the Main Sample 
as described in \citet{Walcher:2014}. There are 3 galaxies from this program in the CALIFA DR3 {, including NGC5947. Those are only ones included in the Pilot Sample that were observed with the final CALIFA configuration and are part of the SDSS imaging survey (a pre-requisite for being distributed in this Data Release). All the three fulfill the Quality Control criteria.}

\subsubsection{Supernova environments}

The extension program on core collapse supernova (CCSN) environments is led by Galbany. 
The galaxies in this program have IDs between 5000 and 5999. 
The project aims to get imaging spectroscopy for low-mass galaxies that hosted type Ib, Ic and II supernovae with 
available light curves. 
This sample increases the completeness of the SN host galaxy sample from the CALIFA Survey presented in 
\citet{Galbany:2014} and 
Galbany et al. (submitted), which has a deficit of CCSNe in low-mass ($\lesssim 10^{10}$ M$\odot$) 
galaxies. 

Objects were selected from the Asiago SN catalog\footnote{http://graspa.oapd.inaf.it/asnc.html} \citep{barbon99} 
following these criteria: a) SN projected galactocentric distance lower than 40 arcsec, in order to cover the 
local SN environment; b) Systemic velocity of the galaxy lower than 9000 km/s ($\sim$z$<0.03$); c) log D$_{25}$ 
(decimal logarithm of the apparent 25 mag/arcmin$^2$ isophotal diameter) lower than 1.12, which corresponds to 
galactic radius lower than 40 arcsec; d) Declination $\delta> 0^{\circ}$; and e) SN light curve publicly or privately available. 

The input sample contains a total of 49 objects of which 14 were observed and included in the CALIFA DR3.

\subsubsection{Compact early-type galaxies}

The extension program on compact early-type galaxies is led by Y{\i}ld{\i}r{\i}m and van den Bosch and includes galaxies with CALIFA IDs between 9000 and 9999. The scientific aim is the characterisation of extremely compact host galaxies of supermassive black holes. Galaxies were selected from the HETMGS \citep[Hobby-Eberly Telescope Massive Galaxy Survey][]{vandenbosch2015} to have: i) large black hole spheres of influence; ii) 2MASS half-light radii smaller than 2 kpc; iii) very high central density as measured by the velocity dispersion over central resolution element ($\sigma$/$1"^2$). Scientific results and further descriptions of the sample can be found in \citet{walsh15} and  \citet{yildirim16}. In those papers, the data are deeper than the standard CALIFA depth, as the exposure times are four times longer. For the sake of uniformity of the DR3, we limit ourselves to the standard CALIFA depth, i.e.~exposure time. Observations are taken with the V500 low-resolution setup only.

Only 7 of the 17 galaxies are included in this data release as the remainder lack SDSS imaging needed for the CALIFA pipeline.

\subsubsection{Other extension programs}

There are two galaxies in DR3 from extension programs that yielded only 
one released cube, and it is therefore not useful to describe these programs in detail.
The galaxy with CALIFA ID number 7001 is CGCG263-044, classified as Sb in NED, is relatively edge-on and classified as a
Type 2 AGN. The galaxy with CALIFA ID number 8000, NGC426, is a massive galaxy, classified as a cD.

\subsection{Properties of the released galaxies}

The physical properties each galaxy in each sample, including its name, CALIFA ID, coordinates, redshift, 
photometry, morphology, and stellar mass are available from the DR3 web page. 

As clearly seen in Figure \ref{fig:morph_mstar}, the morphological and stellar mass distributions of the 
Main Sample are consistent with those of the mother sample. While larger differences 
are seen at low stellar masses, this follows from the low number of galaxies overall in this 
mass range. This test therefore implies that the selection of Main Sample galaxies from the mother sample using target visibility preserves the mother sample statistics. 

\begin{figure}
\resizebox{\hsize}{!}{\includegraphics{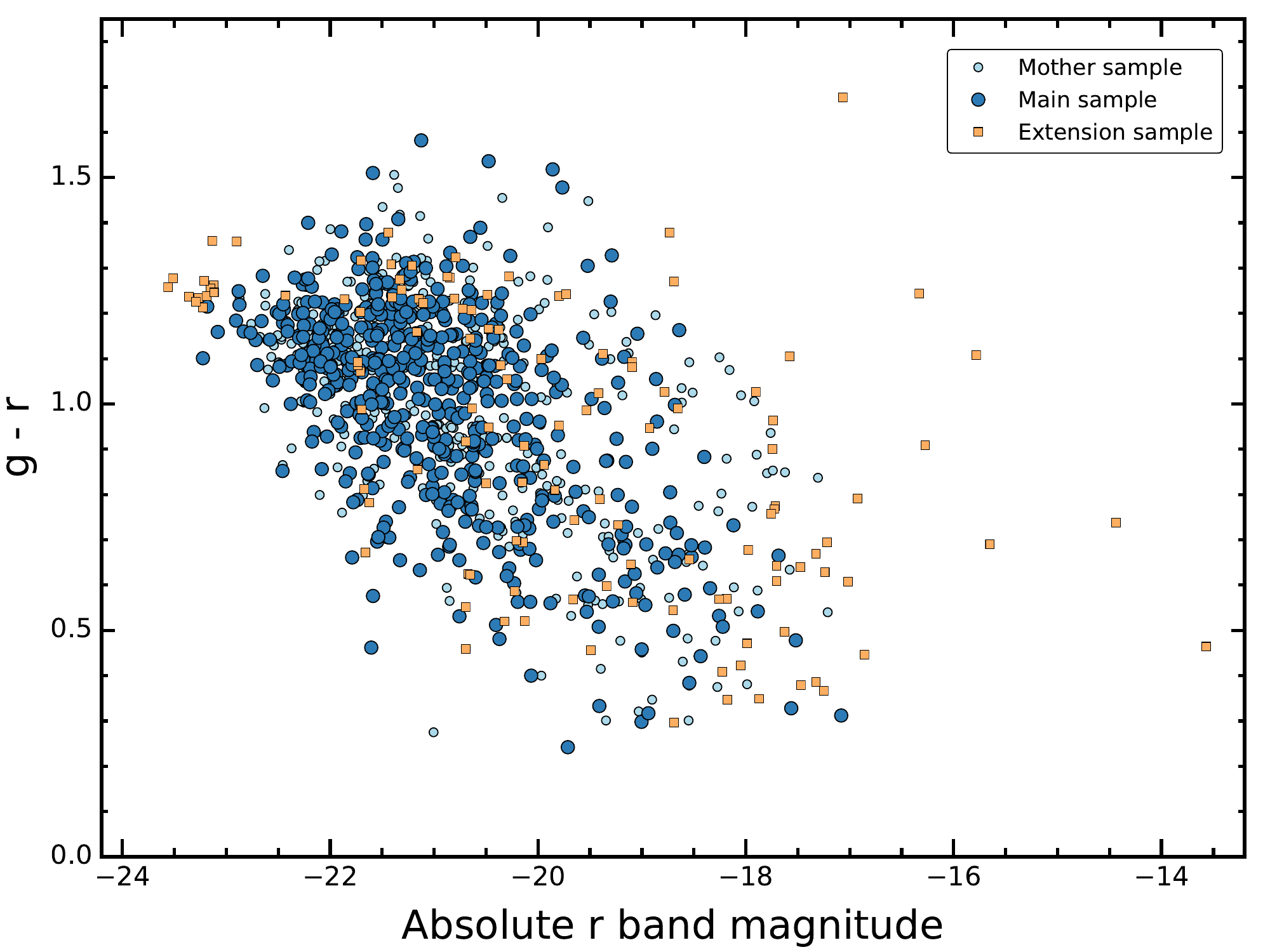}}
  \caption{The distribution of DR3 galaxies in the colour magnitude diagram. }
  \label{fig:col_mag}
\end{figure}

Figure \ref{fig:col_mag} shows the distribution of galaxies in the colour-magnitude diagram. Again, the good coverage 
of the colour-magnitude diagram found for the mother sample is retained for the Main Sample. 
The Extension 
Sample \emph{by design} covers those regions that were not included in the mother sample, i.e.~the bright end 
of the red sequence and dwarf galaxies. 

In \citet{Walcher:2014} we reported on the average spatial coverage of the mother sample galaxies by the IFU as a fraction of 
the Petrosian half light radius (SDSS pipeline quantity PetroR50) $r_{50}$ as computed from the growth curve photometry. About 97\% of all galaxies are 
covered out to at least 2$\times$$r_{50}$. This statement holds for the Main Sample as well. 
Because growth curve photometry is not available for the Extension Sample, we refer here to the SDSS 
Petrosian half light radius, which we will denote as $r_e^p$. The average spatial coverage in terms of $r_{50}$ 
is 4.2 for the Main Sample and 7.9 for the Extension Sample, with the broader coverage of the Extension Sample being driven by the inclusion of dwarf galaxies therein. All Extension Sample galaxies are covered 
out to the SDSS isophotal major axis, as is the case for the Main Sample. 

\begin{figure}
\resizebox{\hsize}{!}{\includegraphics{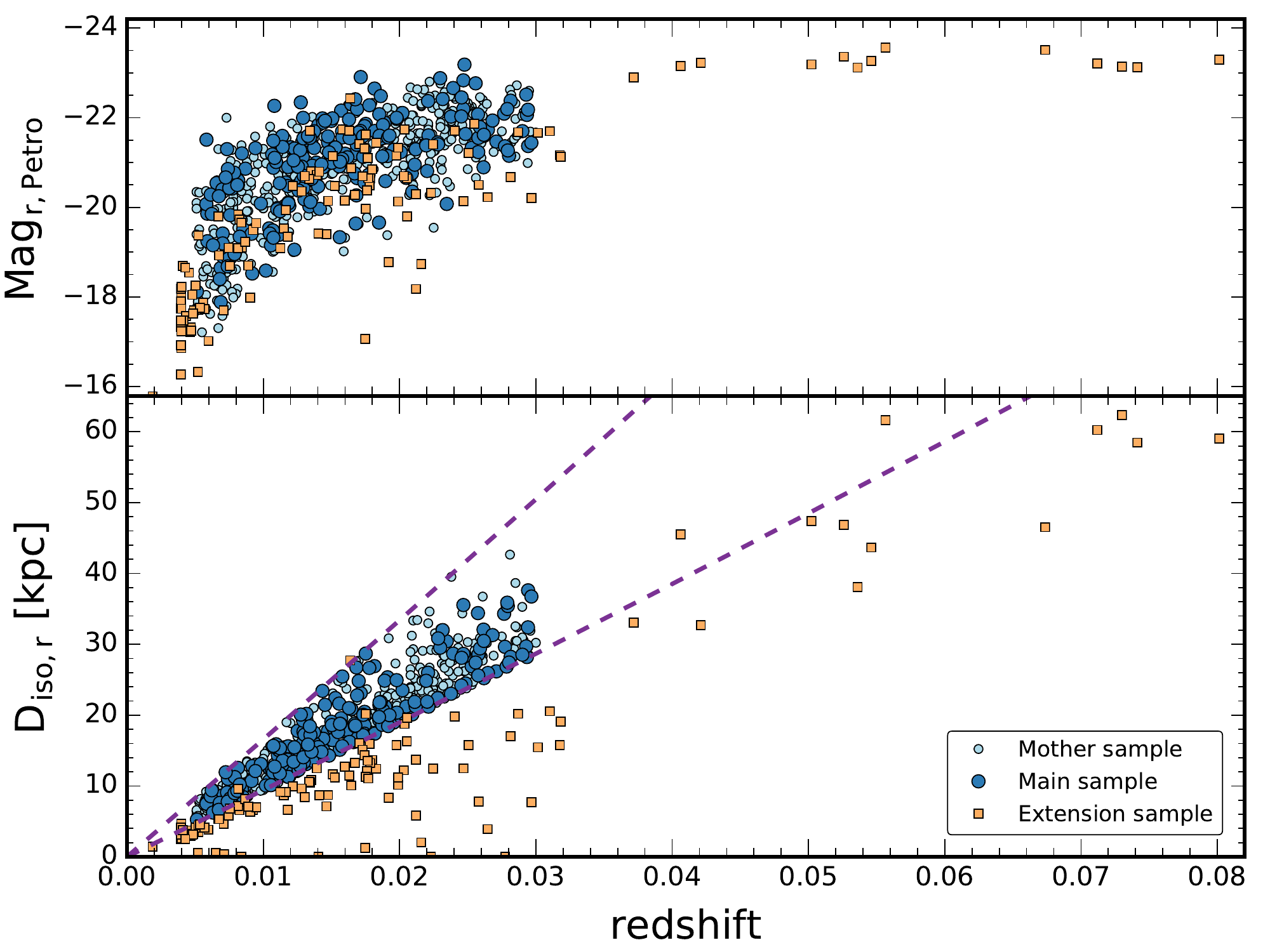}}
  \caption{The distribution of DR3 galaxies in the redshift-magnitude (upper panel) and redshift-size (lower panel) diagrams. }
\label{fig:z_isoA}
\end{figure}

Figure \ref{fig:z_isoA} shows the distribution of galaxies from the Main Sample and Extension Sample in the
redshift-magnitude and redshift-size diagrams. Clearly, galaxies slightly larger than the CALIFA size limit 
are so rare that they are under-represented within the CALIFA volume \emph{independently of their size}. 
Finding them requires an extension to higher redshift. Galaxies smaller than the CALIFA size limit are 
abundant in the local universe. These dwarf 
galaxies were not included in the mother sample \emph{by design} to avoid swamping the Main Sample with them. 
The dwarf galaxies in the Extension Sample somewhat helps to circumvent this self-imposed limitation. 
Still, a dedicated dwarf imaging spectroscopy survey of similar size to CALIFA is missing in the literature. 

\subsection{Luminosity functions of the Main Sample}

\begin{figure}
\resizebox{\hsize}{!}{\includegraphics{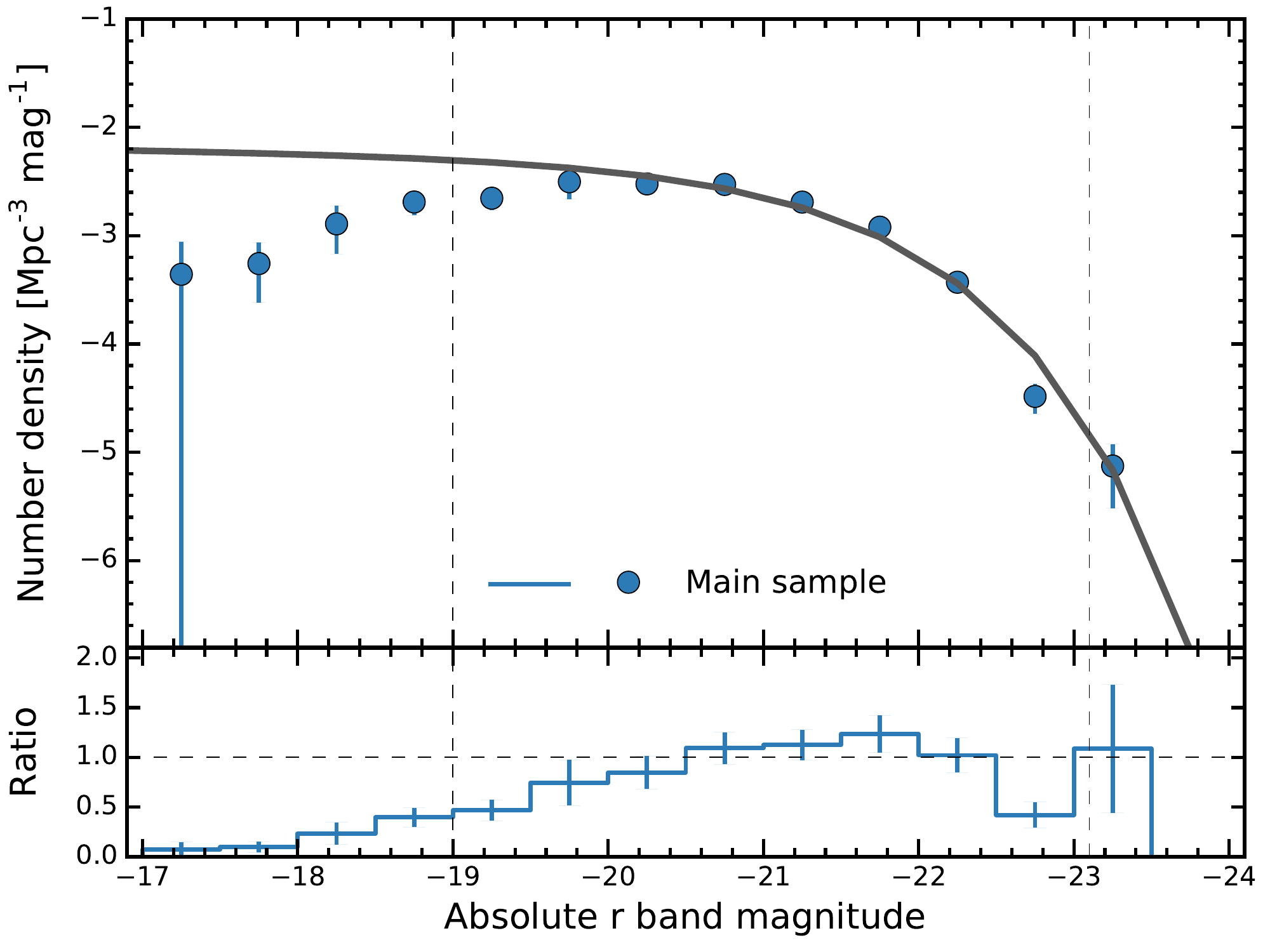}}
 \caption{\emph{Top panel:} $r$-band luminosity functions for the SDSS reference sample from 
 \citet{Blanton:2005} (solid line) and the DR3 Main Sample. 
 Error bars represent Poissonian uncertainties. \emph{Bottom panel:} ratio between the Main Sample and SDSS.}
\label{fig:DR3_LF}
\end{figure}

In \citet{Walcher:2014} we established that the luminosity function of the CALIFA mother sample compares 
well with the reference SDSS sample of \citet{Blanton:2005}. We now investigate whether the same statement 
can be made for the Main Sample. For all technical details on how the luminosity functions are 
obtained and for the derivation of the CALIFA mother sample completeness limit at $M_r \approx -19$ we refer 
the reader to \citet{Walcher:2014}. 

The only difference in the derivation of the luminosity function 
for the Main Sample is the fact that the available volume has been made smaller by a random subselection from the 
mother sample. To correct for this, one needs to multiply the V$_{\mathrm{max}}$ values of the Main Sample galaxies 
with a ratio of the number of galaxies in the sample in use (e.g.~542 in the case of the full Main Sample) divided 
by the number of galaxies in the mother sample (937). The precondition for this simple procedure is that the 
subsample in use can be considered a random subset of the mother sample. 
{ We compared the Mass and Morphology distributions of the Main Sample and 
the Mother Sample using a $\chi^2$ test and found that the probability that they 
were drawn from the same distributions are 98\% and 97\%, respectively. Thus we 
conclude that the Main Sample is a random subselection from the Mother Sample.}

Figure~\ref{fig:DR3_LF} shows the $r$-band luminosity function of the DR3 Main sample compared 
to the Schechter function derived in \citet{Blanton:2005}. Within the statistical uncertainties, the DR3 
Main Sample reproduces the standard luminosity function. { There are two points at the 
limits of our completeness range (at -19.25 and -22.75 in absolute magnitude) that 
seem to deviate more than the 1$\sigma$ range of their errorbars. It is not possible 
to formally decide whether this indicates a (small) issue with our completeness or 
whether this is just as expected from statistics ($\sim30$\% of points should 
lie outside the 1$\sigma$ errorbar).  In any case, a formal $\chi^2$ test indicates a probability of 99.9\% that the two 
functions are identical}. We thus conclude from this section that the Main Sample is representative of the galaxy population 
within the same limits indicated in \citet{Walcher:2014}. This is an important result that allows the 
use of CALIFA data 
to construct galaxy distribution functions in all scientifically useful parameters. One should keep in mind, 
however, that not all mass ranges are equally well sampled statistically. 


\section{Observing strategy and setup overview}\label{sect:observations}

The observing strategy and setup of the CALIFA survey were described in detail in \citet{Sanchez:2012a} 
and have not changed during the survey's six year duration. For the sake of completeness, we provide here 
a very brief summary.

All galaxies were observed using PMAS \citep{Roth:2005} in the PPak configuration 
\citep{Verheijen:2004,Kelz:2006}. The PPak 
science bundle was created to cover a wide area on sky following the requirements of the Disk Mass Survey
\citep{Bershady:2010}, and is now a common-user instrument. The PPak Integral Field Unit (IFU) has a Field of View (FoV) of 74\arcsec\ $\times$ 64\arcsec 
and it contains a total of 382 fibers, distributed in three different groups.  
331 ``science'' fibers are arranged in a hexagonal grid, with each fiber projecting to 2\farcs7 in diameter on the sky. 
The fiber-to-fiber distance is 3\farcs2, which yields a total filling factor of $\sim$60\%. An additional set of 36
fibers located in a ring with radius 72\arcsec\ measure the surrounding sky background. Finally, 15 fibers are connected to the calibration unit.

The goal of CALIFA was to observe every sample galaxy using two different overlapping setups.
The red low-resolution setup (V500; R~$\sim$~850) covers the wavelength range 3745-7500 \AA. The spectra on the CCD are 
affected by internal vignetting within the spectrograph, giving an unvignetted range of 4240-7140 \AA. The blue 
mid-resolution setup (V1200; R~$\sim$~1650) covers the range 3400-4840 \AA, with an unvignetted range of 
3650-4620 \AA. The resolutions quoted are those at the overlapping wavelength range ($\lambda$ $\sim$ 4500 \AA). 
However, since observing in the V1200-setup was more time consuming and required more restrictive weather conditions, not all galaxies were finally observed in both setups. For those that were observed in both setups, the quality control rejected a larger fraction.

A three pointing dithering scheme was used for each object in order to reach a filling factor of 100\% across the 
entire FoV of the science fibers. The dither comprises the following pattern of offsets: (0,0), (-5.22,-4.53), 
and (-5.22,4.53), in arcsec with respect to the nominal position of the galaxy. This pattern comprises a jump of two
inter-fiber distances instead of one to avoid sampling the same area on sky with the same fiber. We thus minimize 
the effects of low transmission fibers in the final dataset.

The exposure times per pointing were fixed, and have been selected to match the S/N requirements of
\citet{Sanchez:2012a}. We carried out V1200 observations during dark nights with an exposure time of 1800 s per
pointing (split into 2 or 3 individual exposures). We obtained V500 observations during grey nights with 900 s per pointing.

For the observations of the Main Sample, target galaxies were randomly selected from the mother sample. The strategy for observing Extension Sample galaxies was more varied, and depended on the extension program.  
Altogether, 685 galaxies were observed during the last 6 years, 
making use of 237 clear equivalent nights distributed between the 6th of June 2010 and the 16th of April 2016.
The number 237 is obtained from 176 totally clear nights, plus observations distributed in partially clear 
nights or among other service programs. Data cubes for 667 of these are in DR3, following strict quality control procedures. 

In the following section, we describe the improvements to the CALIFA data reduction pipeline 
used to produce the DR3 data. 

\section{Data processing and error propagation}\label{sect:data_processing}

\subsection{Overview of the reduction scheme}\label{sect:reducview}

The CALIFA data reduction was performed by a semi-automatic pipeline that follows the procedures for the reduction 
of fiber-fed IFS data described in \citet{sanchez06a}. The CALIFA data reduction pipeline was first (until 
V1.2) developed in {\tt Perl} and named {\sc R3D}\footnote{\url{http://www.astroscu.unam.mx/~sfsanchez/r3d/}}. 
It was then rewritten using a {\tt Python}-based core in the {\sc Py3D} package 
\citep[V1.3c and V1.5][]{Husemann:2013,rgb15}. The current pipeline version, V2.2, has now fully replaced the 
original scheme and uses {\tt Python} for the full process.

The reduction process comprised the following steps: i) The four different FITS files created by the
amplifiers of the detector were re-arranged into a single frame, which is then bias subtracted. Cosmic rays were 
removed and masked according to \citet{Husemann:2012}; ii) Relative offsets in the tracing due to flexure were 
estimated with respect to the continuum and arc-lamp calibration frames and the wavelength solution was applied to
each individual science frame; iii) The stray-light map was reconstructed using the gaps between the fiber 
traces\footnote{See \citet {Kelz:2006} for a description of the spatial arrangement of the fibers on the 
pseudo-slit and CCD.} and subtracted from the calibration and science exposures; iv) An optimal extraction 
algorithm \citep{Horne:1986} was used to extract the spectra based on measurements of the positions and widths 
derived from the continuum lamp. { The optimal extraction relies on a good characterization of the shape of the spectra along the cross-dispersion axis. In theory it is required to make a measurement considering both the Gaussian core and the Lorentzian wings. However, in practical terms the adopted procedure to subtract the stray-light removes most of the Lorentzian wings, and therefore a pure Gaussian function is a good representation of the shape of the spectra}; v) The extracted flux for each pixel in the dispersion direction was stored in 
a row-stacked-spectrum (RSS) file; vi) The spectra were resampled to a linear grid in wavelength using the 
wavelength solution and resolution obtained from the HeHgCd calibration lamp exposure taking for each pointing, 
taking into account possible flexure offsets in the dispersion axis within each pointing; vii) In the same step, 
the spectral resolutions were homogenized to a common value along the dispersion axis using an adaptive Gaussian 
convolution kernel.

The Poisson shot noise, the read-out noise, and bad pixel masks were propagated in the reduction process. For 
the wavelength solution, errors were analytically propagated during the Gaussian convolution and a Monte Carlo 
approach was used to estimate the noise vector after the spline resampling of the spectra. Fiber-to-fiber 
transmission throughput was corrected using an RSS master fiberflat created from sky exposures taken 
during twilight on all survey observing nights.

Flux calibration was performed using the procedure described in \citet{rgb15}. In essence we performed a 
dedicated calibration program, re-observing two dozen CALIFA ETGs chosen as secondary calibrators 
and a set of the standard stars with the PMAS Lens-Array (LArr) (Husemann et al., in prep.). This IFU covers a 
continuous $16\arcsec\times16\arcsec$ FoV which produces highly accurate spectrophotrometric spectra.
Comparing the photometry of the calibrated LArr data with aperture-matched SDSS photometry in the $g$ 
and $r$ bands, the absolute spectrophotometric accuracy of our standard galaxies is $<$0.03\,mag. 

During the last 4 years of the survey (2012-2015) we were observing the ETG calibration galaxies regularly, 
and updating the master sensitivity curve of the instrument/setup created as described in \citet{rgb15}. 
We adopted that sensitivity curve to perform the spectrophotometric calibration for DR3. For each 
particular pointing the flux calibration was performed by correcting for the atmospheric extinction using 
the mean observatory curve 
presented by \citet{Sanchez:2007}, and using the extinction ($A_{V}$) provided by the Calar Alto Vistual 
EXtinction monitor (CAVEX) measured at the moment of the observations. When the CAVEX was not operating 
($\sim$15\% of the time) we adopted the average extinction at the observatory ($A_{V}\sim0.15$ mag). 
Most of the remaining systematic effects in the spectrophometric uncertainty for CALIFA are driven by the 
uncertainties in the wavelength-dependent atmospheric extinction at the time of each observation, that is 
not properly monitored at the observatory.

The science spectra corresponding to the three dithered exposures were combined into a single frame of 993 spectra,
following the registration procedure described in \citet{rgb15}. In summary, we computed the flux 
corresponding to the 331 apertures of the fibers for each pointing from sky-subtracted SDSS DR7 images in 
the bands covering the wavelength of our observation. The apertures were shifted in right ascension and 
declination over a search 
box around the nominal coordinates of the pointing and the best registration was found on the basis of a 
$\chi^2$ comparison between the SDSS aperture-matched fluxes and those derived from the RSS spectra themselves. 
This provided us with accurate astrometry for each pointing (with a typical error of $\sim$0.2\arcsec), and 
a flux recalibration pointing by pointing. This recalibration anchors the absolute CALIFA spectrophotometry 
to that of the SDSS DR7.

After the Galactic extinction correction \citep{schlegel98,cardelli89}, the RSS was ready for the spatial 
rearranging of the fibers and creation of the datacube. We used a flux-conserving inverse-distance weighting 
scheme to reconstruct a spatial image with a sampling of 1\arcsec\ using the same parameters as described in 
\citet{rgb15}. This scheme is now adopted by other IFU surveys such as MaNGA \citep{manga}, 
as described by Law et al. (in prep.). First, we reconstructed the datacube and estimated the differential 
atmospheric refraction (DAR) offset. In a second step, we reconstructed the cube again but shifting the position 
of the fiber at each wavelength against the regular grid according to the DAR offset measured in the first 
reconstruction. This two-stage iteration avoids one resampling step, which is important for accurate error 
propagation. 

\begin{figure*}
\resizebox{\hsize}{!}{\includegraphics{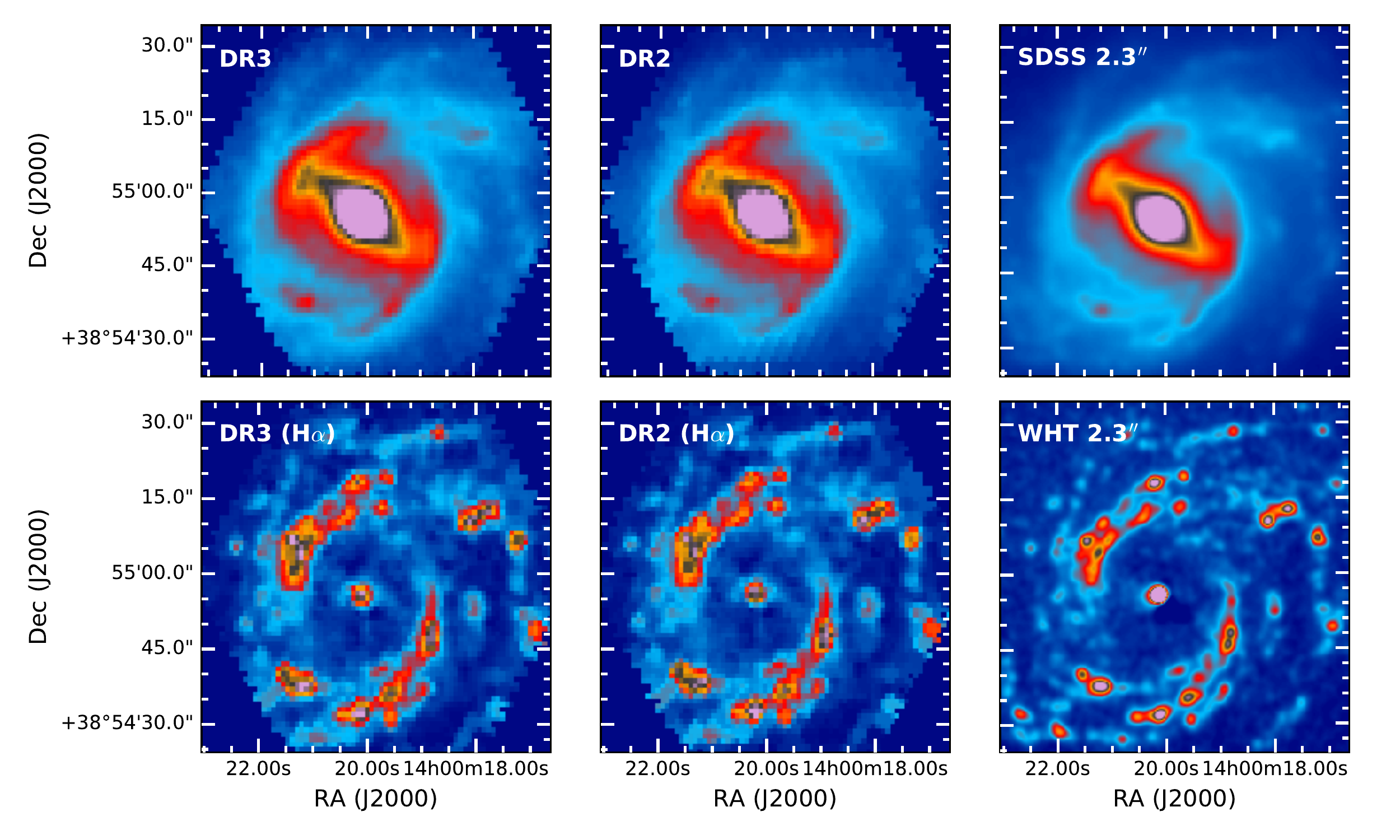}}
\caption{{\it Top panel:} Comparison between $g$-band images extracted from CALIFA datacubes of the galaxy NGC~5406 (ID=684) created using 
the V2.2 (DR3) and V1.5 (DR2) pipelines, together with the corresponding 
SDSS image convolved with a Gaussian function to match the spatial resolution of the CALIFA datacubes. 
{\it Bottom panel:} Similar comparison between H$\alpha$ images extracted from the same datacubes using 
the Pipe3D pipeline, and a resolution-matched H$\alpha$ image taken with the 4.2m William Herschel Telescope 
(Roque de los Muchachos Observatory, La Palma, Spain), using the AUXCAM detector 
(S\'anchez-Menguiano et al., in prep.). The FoV has been reduced to match that of CALIFA.}
  \label{fig:DR3_spatial}
\end{figure*}

\subsection{Improvements to the CALIFA data reduction scheme}\label{sect:pipeline}

The main improvements to the pipeline when going from V1.5 (DR2) to V2.2 (DR3) are: i) a new version of the 
high-level scripting code that handles the night-by-night reduction; ii) an improved version of the COMBO datacubes, 
i.e., the cubes created by combining the datasets from the V500 and V1200 setups \citep[already used in different 
science studies, e.g.][]{rosa15a,rosa16}; iii) fully automatic verification of the registration and astrometry 
process; iv) a second-order correction of the datacube reconstruction based on a comparison with the SDSS images.

As indicated before, the new high-level routines that handle the reduction are now written in Python 2.7. 
Like in the previous version the low-level routines 
of the reduction are based on the {\sc Py3D} package, and therefore the reduction sequence 
and detailed routines have not changed since V1.5 \citep{rgb15}. The use of Python for the 
high-level routines provides three basic improvements: i) portability, i.e.~the pipeline can be easily installed 
on any workstation, which is more user friendly; ii) maintenance, i.e.~future updates to the pipeline will be 
easier; and iii) speed, thanks to the multi-processing module. Basically all objects of the same type 
(science objects, calibration stars, etc.) observed during a night can be processed in parallel.

The pipeline creates a new version of the COMBO cubes by taking into account the data from both the V500 and V1200 setups 
in the overlapping areas of the spectrum. These COMBO cubes are created to solve the problem of vignetting affecting both setups at the edge of their spectral wavelength ranges \citep[Section 4.2 of][]{rgb15}. The COMBO cubes span an unvignetted wavelength range of 3700-7140 \AA.

In the previous version (not distributed in DR2 and DR3), the COMBO cubes were created by glueing the 
data from both datasets together at a cut-off wavelength. Specifically, the 
resolution of the V1200 cube was degraded to that of the V500 cube and the cubes were recentered. Then, the 
resolution-matched V1200 cube was used below a cut-off wavelength of 4500{\AA}. Above that wavelength the V500 cube 
was used. In order to avoid low-level spectrophotometric mismatches the V1200 spectra were re-scaled to the V500 
ones by the average of the ratio between both datasets in the overlapping wavelength range.

The COMBO cubes in DR3 were created by using the spectra from both setups in the overlapping region simultaneously. 
Like in the previous case, the V1200 data were spatially recentred, flux rescaled and degraded in resolution to match 
the V500 data. Then the COMBO cube was created by averaging the spectra corresponding to each dataset in the overlap
region, weighted by the inverse of the cube error. For the remaining wavelength range the COMBO cube consists of only 
the V1200 or the V500 spectra. This new procedure improves the S/N in the overlap region of the COMBO cubes. 

The current spatial registration scheme is the same as the one described in \citet{rgb15}. It is known that
this registration process fails in some cases, particularly in low surface brightness and/or edge-on galaxies 
or in the presence of bright foreground field stars. These failures happened more frequently in the V1200 setup, given 
its lower S/N on average compared to that of the V500 setup. The current V2.2 pipeline automatically discards the 
registration procedure when the minimum $\chi^2$ is higher than a given threshold. Then it only applies 
a global flux re-scaling as described in \citet{Husemann:2013}, relying on the nominal offsets provided by the 
telescope for the World Coordinate System \citep[WCS][]{greisen02}. A boolean header keyword ({\tt REGISTER}) is 
added to the datacube to indicate whether the cube has been fully registered or not. { In \cite{rgb15} we found that the astrometric solution has an accuracy better than 3$\arcsec$ for $\sim$93\% of the targets. We repeated the analysis for the new data set and we find that there is a better precision in our astrometry, with a standard deviation in the offset with respect to the SDSS one of $\sim$0.6$\arcsec$ in both RA and DEC. However, we have a systematic offset of $-$0.6$\arcsec$ in both directions (e.g., $RA_{\rm CALIFA}-RA_{\rm SDSS}$), whose source is still unclear.} 

Finally, a second-order correction was applied to the CALIFA datacubes to match their spectrophotometry as much as 
possible to that of the SDSS images. This procedure followed a two-step process. First, a second-order correction 
to the WCS astrometry of the cubes was obtained.
Sky-subtracted SDSS DR7 $g$- (for the V1200) and $r$- (for the V500) 
band images were downsampled to 1$\arcsec$/pixel. The corresponding images using the proper filter response curve 
were created from the CALIFA datacubes. The synthetic CALIFA image and SDSS image were registered using a Discrete 
Fourier Transform and the offsets are updated in the CALIFA image and cube headers. We then convolved the SDSS 
image with a sequence of circular Moffat kernels,
\begin{equation}
 I(r) = I_0 \left[1+\left(\frac{r}{\alpha}\right)^2\right]^{-\beta},\label{eq:Moffat}
\end{equation}
varying $\alpha$ and $\beta$, i.e.~effectively varying the Full Width Half Maximum (FWHM), which depends on both parameters,
\begin{equation}
\mathrm{FWHM} = 2\alpha\sqrt{2^\frac{1}{\beta}-1}.\label{eq:FWHM}
\end{equation}
Each convolved SDSS image was then compared to the corresponding 
reconstructed CALIFA images and the best parameters are selected by $\chi^2$ minimization. This 
procedure provides the differential spatial resolution between the SDSS images and the CALIFA cubes. Taking 
into account the FWHM of the Point Spread Function (PSF) for the SDSS images (that on average is
$\sim$1.5$\arcsec$) we obtained a 
first order estimate of the FWHM of the CALIFA PSF for each cube. This is stored in the header keyword 
{\tt FWHM} in units of arcsec.

Once the convolved SDSS image that best reproduces the one reconstructed from CALIFA was obtained, 
we computed their ratio. 
This ratio, called {\it SDSSflat}, is a 2D map with a mean value of one and a scatter of a 
few percent across the FoV. The final correction was applied by multiplying the fluxes and variances of the 
data with this 2D map, changing the photometric absolute scale in each spaxel, without affecting the shape 
of the spectra. The flux level of the integrated spectrum for each datacube changed by less than a few percent 
both in absolute and relative terms (from blue to red), due to the different relative contribution of each 
individual spaxel to the sum.

The {\it SDSSflat} is stored in an additional extension in the FITS files named {\tt FLAT}. In some cases, in 
particular when there are registration issues and/or very bright field-stars, the procedure fails. This was 
easily identified during the QC process since the distribution of values within SDSSflat was not symmetric, 
was not centered on $\sim$1, and its application to the cube modified the shape of the integrated spectrum 
by more than the expected spectrophotometric accuracy ($\sim$3\%). In those cases we have preferred not 
to apply the correction. Whether this correction is applied or not is indicated by a header keyword 
({\tt FLAT\_SDSS}), that is set to {\tt true} or {\tt false}.

Figure \ref{fig:DR3_spatial} illustrates the improvements and similarities between V2.2 (DR3) and V1.5 (DR2) 
of the data reduction by comparing i) the $g$-band images extracted from both datacubes of the galaxy 
NGC~5406 (ID=648) with the corresponding resolution-matched $g$-band SDSS image, and ii) the $H\alpha$ 
intensity map extracted from both datacubes using the {\tt Pipe3D} pipeline \citep{sanchez2016b}, 
together with a narrow-band image centered on the same emission line. This last image was taken with the 
4.2m William Herschel Telescope (Roque de los Muchachos Observatory, La Palma, Spain) using the AUXCAM 
detector (S\'anchez-Menguiano et al., in prep.). The effects of the {\it SDSSflat} are visible through the absence 
of patchy structures in the broad-band images from V2.2, evident in the V1.5 image. Finally, the spatial 
resolution is not affected, as clearly seen in the similarities between the three H$\alpha$ images. 

\subsection{Characterization of spatially correlated noise}\label{sect:correlation}

Like in the case of V1.3c and V1.5 the interpolation procedure used to obtain a regular grid implies that 
the output pixels in the final datacube are not independent of each other. The Gaussian interpolation method 
distributes the flux from a given fiber to several pixels, which are combined with neighboring pixels 
within a certain radius \citep[see Section 4 of][]{rgb15}. This causes the noise in the adjacent pixels to be 
spatially correlated. Recall that even in the case that there is no interpolation of the RSS files, all 
spectra are correlated at some level due to their projection on the detector. This correlation is stronger 
in adjacent spectra at the detector level, that are not necessarily adjacent in the focal plane of the telescope 
\citep{Kelz:2006}. This correlation implies that a measurement of the noise in a stacked spectrum of $N$ 
pixels will be underestimated. Characterizing this effect is essential for estimating the statistical errors 
when spectra in datacubes are coadded. 

\begin{figure}
\resizebox{\hsize}{!}{\includegraphics{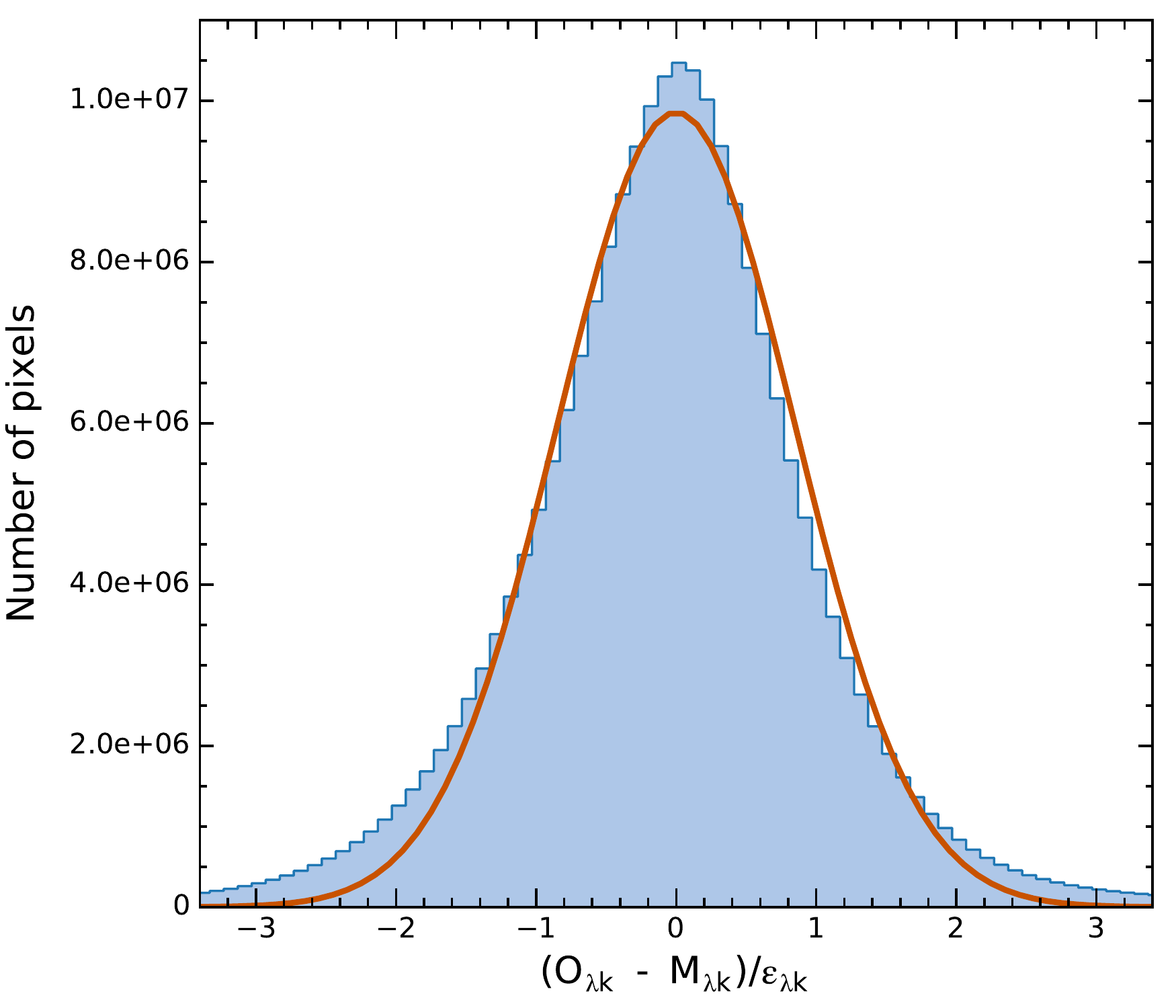}}
  \caption{Histogram of the reduced residuals $(O_{\lambda,k} - M_{\lambda,k}) / \epsilon_{\lambda,k}$ for 
     all wavelengths, all bins ($k$) and all galaxies in DR3 (433400381 points in total). The orange line shows the best Gaussian fit to the sample.}
  \label{fig:DR3_hist_error}
\end{figure}

Following \citet{Husemann:2013} and \citet{rgb15} we checked that the error spectra derived from the pipeline 
for individual spaxels are reliable. Spectral fitting analysis can provide an approximate assessment of the accuracy 
of the error spectra. In Figure~\ref{fig:DR3_hist_error} we update Figure~9 of \citet{Husemann:2013} and Figure~10 of
\citet{rgb15} for DR3 data. The plot shows the histogram of residuals, i.e., the difference between the observed
($O_{\lambda}$) and synthetic ($M_{\lambda}$) spectra obtained with {\sc Pipe3D} in units of the corresponding 
error $\epsilon_{\lambda}$ (details of the fitting procedures can be found in Section \ref{sect:specphot_cal}). The
distribution is very well described by a Gaussian centered at 0.01 with $\sigma = 0.83$, only slightly lower than 
expected if residuals are purely due to uncorrelated noise. 

\begin{figure}
  \resizebox{\hsize}{!}{\includegraphics{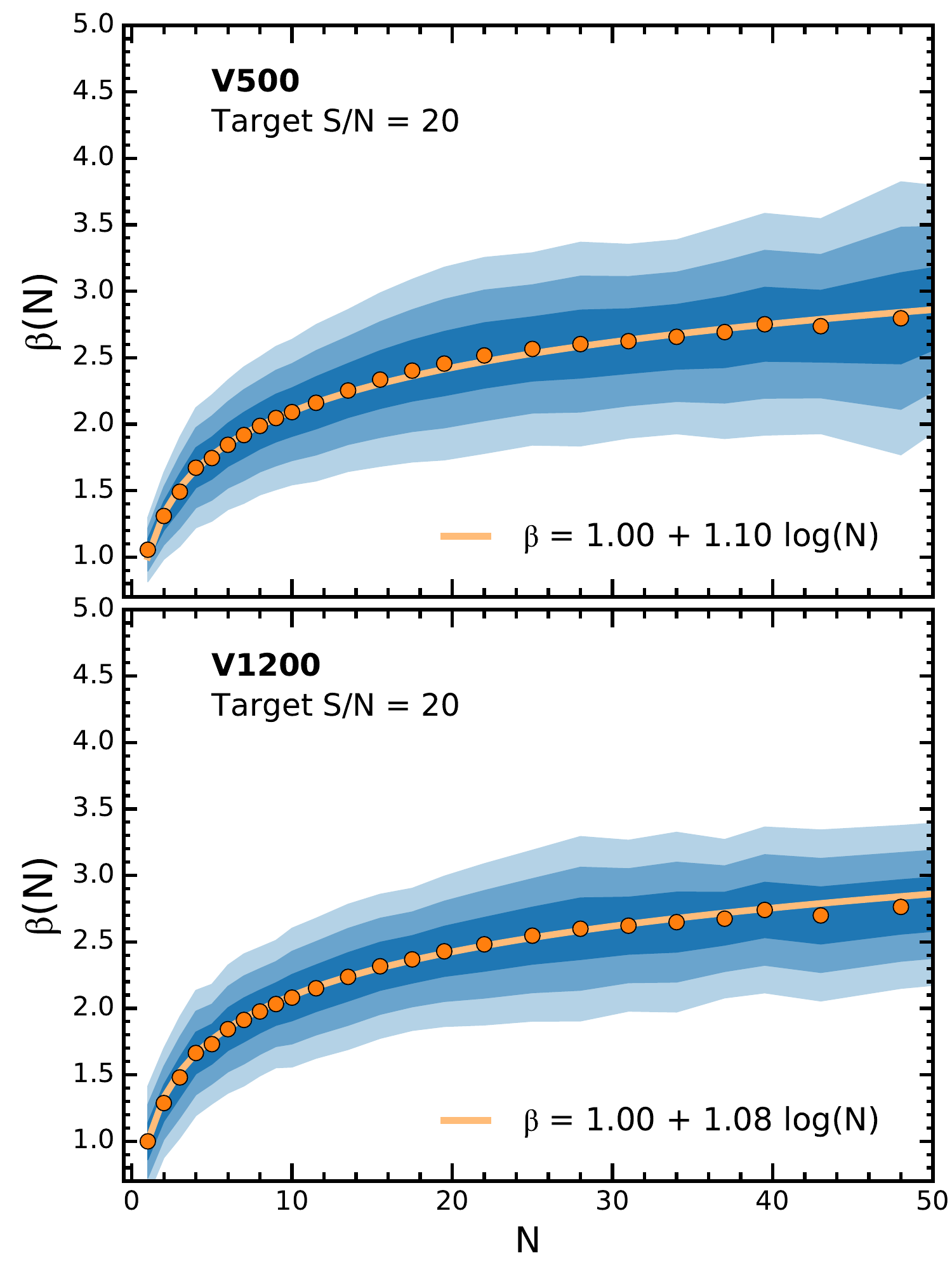}}
  \caption{Noise correlation ratio $\beta$ (ratio of the real estimated error to the analytically 
    propagated error) as a function of number of spaxels per bin for all the V500 \emph{(top panel)} 
    and V1200 \emph{(bottom panel)} data of DR3 at a target S/N of 20. Shaded areas mark the 
    1-$\sigma$, 2-$\sigma$,  
    and 3-$\sigma$ levels. The orange lines represent the best fitting logarithmic function with a slope 
    $\alpha = 1.10$ and $\alpha = 1.08$, respectively.}
  \label{fig:DR3_noise_correlation}
\end{figure}

The correlated noise can be taken into account by providing the spatial covariance \citep{Sharp:2014}. 
However, like in DR1 and DR2, a more practical approach consists of using the datacubes to introduce the noise 
correlation ratio as a function of the number of pixels $\beta (N)$. $\beta$ is the ratio of the ``real'' or 
measured error to the analytically propagated error of the binned spectra as a function of bin size. 
To calculate $\beta$ We used the Voronoi adaptive binning method 
\citep[implemented for optical IFS data by][]{Cappellari:2003} with a target S/N of 20 to obtain a sample 
of coadded spaxels covering different numbers of spaxels. We removed individual spaxels with S/N $<$ 5 from 
the analysis, and coadded bins with areas larger than 60 spaxels. The ``real'' noise was then obtained from 
the detrended standard deviation in certain defined wavelength windows (see Section \ref{sect:depth}).), 
The results obtained for all DR3 datacubes, shown in Figure \ref{fig:DR3_noise_correlation}, 
can be described well by the following logarithmic function: 
\begin{equation}
 \beta (N) = 1+\alpha\log N,\label{eq:correlation}
\end{equation}
where $N$ is the number of spaxels per bin.

The values for the slope $\alpha$ are very similar in both setups, with a value of <1.10 for V500 and 1.08 
for V1200, with errors in the estimation of the slope around 0.01. Both of them agree well with the observed 
distribution within one sigma. The slope is lower 
than the DR1 value (mean $\sim$ 1.4), but very similar to the value reported for DR2. This indicates that 
the noise in DR3 and DR2 datacubes is less correlated than that in the DR1 datacubes. This is expected 
since we changed the parameters in the interpolation and the registration procedure from V1.3c to V1.5 
and V2.2. Detailed instructions on how to estimate the coadded error spectrum are given in Appendix A 
of \citet{rgb15}.


\begin{table*}
\centering
\caption{CALIFA FITS file structure}
\label{tab:HDUs}
\begin{tabular}{cccc}\hline\hline
HDU & Extension name & Format & Content\\\hline
0 & Primary & 32-bit float & flux density in units of $10^{-16}\,\mathrm{erg}\,\mathrm{s}^{-1}\,\mathrm{cm}^{-2}\,\mathrm{\AA}^{-1}$\\
1 & ERROR & 32-bit float & $1\sigma$ error on the flux density\\
2 & ERRWEIGHT & 32-bit float & error weighting factor\\
3 & BADPIX &   8-bit integer & bad pixel flags (1=bad, 0=good) \\
4 & FIBCOVER &  8-bit integer & number of fibers used to fill each spaxel\\ 
5 & FLAT &  32-bit float & SDSSflat correction to the interpolation \\ \hline
\end{tabular}
\end{table*}

\section{CALIFA data format and characteristics}\label{sect:data_format}

The CALIFA data are stored and distributed as 3D data cubes in the standard  binary FITS 
format. Each FITS file consists of several Header Data Units (HDUs). These HDUs contain, in order within each FITS 
file,  (1) the measured flux densities, corrected for Galactic extinction as described in \citet{Sanchez:2012a}, in 
units of $10^{-16}\,\mathrm{erg}\,\mathrm{s}^{-1}\,\mathrm{cm}^{-2}\,\mathrm{\AA}^{-1}$ (primary datacube), (2) the
associated errors, (3) the error weighting factors, (4) the bad pixel flags, (5) the fiber coverage, and 
(6) the SDSSflat correction to the interpolation scheme (see also Table~\ref{tab:HDUs}). This last HDU 
was absent in DR1 and DR2, as explained in 
Section \ref{sect:pipeline}. The remaining extensions were explained in detail in \citet{Husemann:2013} and \citet{rgb15}.

The first two axes of the cubes correspond to the spatial dimension along right ascension and declination with a
$1\arcsec\times1\arcsec$ sampling. The third dimension represents the wavelength and is linearly sampled.
Table \ref{tab:cube_dimension} summarizes the dimensions of each datacube ($N_\alpha$, $N_\delta$, and $N_\lambda$), 
as well as the spectral sampling ($d_\lambda$) and resolution ($\delta_\lambda$).

\subsection{Error and weight datacubes}

The $1\sigma$ uncertainty of each pixel as formally propagated by the pipeline can be found in the first FITS extension.
Section \ref{sect:correlation} discusses the accuracy of the uncertainties and their correlation properties. This is
important when CALIFA data are spatially binned, and an empirical function is provided to account for the 
correlation effect. 
The second FITS extension (ERRWEIGHT) stores the error scaling factor for each pixel in the limiting case 
that all valid spaxels of the cube would be coadded \citep[see also appendix of][]{rgb15}. In the case of bad pixels, 
we assign an error value that is roughly ten orders of magnitude higher than the typical value. 

\begin{table}
\caption{Dimension and sampling of CALIFA datacubes}
\label{tab:cube_dimension}
\begin{tabular}{lccccccc}\hline\hline
\small{Setup} & \small{$N_\alpha$}\tablefootmark{a} & \small{$N_\delta$}\tablefootmark{a} & \small{$N_\lambda$}\tablefootmark{a} & \small{$\lambda_\mathrm{start}$}\tablefootmark{b} & \small{$\lambda_\mathrm{end}$}\tablefootmark{c} & \small{$d_\lambda$}\tablefootmark{d} & \small{$\delta_\lambda$}\tablefootmark{e} \\\hline
\small{V500}  &  \small{78}    & \small{73}    &  \small{1877}   & \small{3749\AA}     & \small{7501\AA}  & \small{2.0\AA}  & \small{6.0\AA}    \\
\small{V1200} &  \small{78} & \small{73} &  \small{1701} & \small{3650\AA} & \small{4840\AA} & \small{0.7\AA} & \small{2.3\AA}  \\
\small{COMB}  &  \small{78}    & \small{73}    &  \small{1901}   & \small{3701\AA}     & \small{7501\AA}  & \small{2.0\AA}  & \small{6.0\AA}    \\\hline
\end{tabular}
\tablefoot{\tablefoottext{a}{Number of pixels in each dimension.}
\tablefoottext{b}{Wavelength of the first pixel in wavelength direction.}
\tablefoottext{c}{Wavelength of the last pixel in wavelength direction.}
\tablefoottext{d}{Wavelength sampling per pixel.}
\tablefoottext{e}{Homogenized spectral resolution (FWHM) over the entire wavelength range.}
}
\end{table}

\subsection{Bad pixel datacubes}

Bad pixel datacubes are stored in the third FITS extension (BADPIX). This information, in combination with the 
error vector, is essential to properly account for potential problems in each of the pixels. Pixels 
with flag~=~1 report the absence of sufficient information in the raw data due to cosmic rays, bad CCD columns, or 
the effect of vignetting. { They comprise a 4.2\% of the total spaxels in the final datacubes. } The vignetting effect imprints a characteristic inhomogeneous pattern across the 
FoV on the bad pixels vector. More details can be found in Figure 11 of \citet{Husemann:2013}.
These bad pixels have been interpolated over and we strongly suggest not to use them for any science analysis. 

Finally, the uncovered corners of the hexagonal PPak FoV are filled with zeros and flagged as bad pixels for 
consistency. The residuals of bright night-sky emission lines are not flagged as bad pixels. 

\subsection{Fiber coverage datacubes}\label{sect:fcover}

The inverse-distance weighting that is used to reconstruct the data cubes means that several fibers contribute 
to each spaxel for most of the spaxels. As explained in \citet{rgb15} we minimized the maximum distance of fibers 
that can contribute to the flux of a given spaxel to improve spatial resolution. However, the number of contributing 
fibres for spaxels at the edge of the hexagon defined by the dither pattern is lower than for spaxels inside 
the hexagon. As a compromise between improved spatial resolution and avoidance of information loss in the outer 
parts of the hexagon since pipeline V1.5, we also reduced the minimum number of fibers that contribute to each 
spaxel to one. This extension contains the information on the number of fibers used to compute the flux in each spaxel. 

\subsection{SDSS flat-fielding image}\label{sect:SDSSflat}

Pipeline V2.2 has introduced a second-order correction to the interpolation scheme that preserves the 
spectrophotometry at the spatial resolution of our data (see Section \ref{sect:pipeline}). The final correction is 
a multiplicative term that is stored as a 2D image/map in a new HDU (FLAT). The correction can be easily undone 
by the user by dividing the {\tt Primary} and the {\tt ERROR} datacubes by the content of this HDU without 
altering any of the other properties of the data cubes. This HDU is present only for those galaxies where 
the SDSSflat correction is applied.

\subsection{FITS header information}\label{sect:fheader}

The FITS header contains the standard keywords that encode the information required to transform the pixel-space 
coordinates into sky and wavelength-space coordinates, following the WCS.
Each CALIFA datacube contains the full FITS header information of all raw frames from which it was created. 
Information regarding observing and instrumental conditions such as sky brightness, flexure offsets, Galactic 
extinction or approximate limiting magnitude is also kept in the FITS header of each datacube. See Section 4.3 
and Table 4 of \citet{Husemann:2013} and Section 5.4 of \citet{rgb15} for nomenclature and a summary of the main 
header keywords and their meaning. 

The most important new keyword added in DR3 datacubes is ``FLAT\_SDSS'', which takes a boolean value. It indicates 
whether or not the SDSS flat correction has been applied \ref{sect:pipeline}. This information is also included in the electronic tables describing the quality of the data.


\section{Data Quality}\label{sect:QC}

The present third and final CALIFA data release (DR3) provides 
science-ready data for a sample of 646 galaxies observed in the V500 setup,
484 in V1200, and 446 combined ``COMBO" cubes. As for the previous data releases, all datacubes
have been checked according to a QC protocol.
The DR3 QC protocol is similar in many respects to the DR2-QC. However, 
some modifications were introduced, which are highlighted in this section.
The end products of the QC procedure are tables of flags that indicate 
the quality of the released data products:
the observing conditions (denoted by the \textsc{obs} prefix), the instrumental
performance and effectiveness of the data reduction (\textsc{red}),
and the accuracy and quality of the final calibrated data products (\textsc{cal}).
QC assessments are based on measured parameters extracted by the
pipeline at different stages of the reduction procedure and on
visual checks of spatially-integrated spectra and wavelength-integrated 
synthetic images.

Each flag can have one of the following values:
\begin{description}
\renewcommand{\labelitemi}{$\bullet$}
\item[$\bullet$] $-1 =$~ undefined
\item[$\bullet$] $~0 =$~ good quality -- \textsc{OK}
\item[$\bullet$] $~1 =$~ minor issues that do not significantly affect the quality -- \textsc{warning}
\item[$\bullet$] $~2 =$~ significant issues affecting the quality -- \textsc{bad}
\end{description}

Flags depending on measured parameters are assigned by checking
against thresholds, as detailed below and summarized in Tables 
\ref{tab:QCflags_def_V500}, \ref{tab:QCflags_def_V1200}
and \ref{tab:QCflags_def_COMB}. { Below each flag name, in the table we also report the percentage
of cubes with \textsc{OK}, \textsc{warning}, \textsc{bad}, and undefined value, in order to
provide a quantitative assessment of the impact of each flag on the QC.}
The thresholds are determined starting from the
actual parameter distribution, so as to flag obvious outliers, by
comparison with the nominal quality requirements of the survey,
and by checking the impact of exceeding such thresholds on the
accuracy of the wavelength and spectrophotometric calibration. Visual checks for each datacube are 
performed by three members of the collaboration. The median of the three independent assessments 
is taken as the corresponding QC flag.
For flags that combine visual classifications and measured
parameters, partial flags are created independently and the worst value is retained as the final flag.
The tables of QC flags, along with the relevant QC parameters,
are available on the DR3 website.

In naming the QC parameters, we adopt the following convention: 
the first part is the category prefix (\textsc{obs}, 
\textsc{red} or \textsc{cal}), followed by a measured parameter, 
and sometimes a final suffix indicating the statistics applied to 
combine the parameter as measured in different observations/pointings/fibers 
(i.e.~\textsc{mean}, \textsc{min}, \textsc{max}, \textsc{rms}).

The QC of the V500 and V1200 setups is based on the same set of parameters
and visual checks, except for the parameters and flags related to the spectrophotometric
comparison with the SDSS (see below and section \ref{subsect:QC_cal}), 
which only applies to the V500 setup.
The COMBO cubes inherit all the flags from the corresponding V500 and V1200
"parent" cubes. In addition they are visually inspected to make sure that
the combination process did not introduce any defects or artifacts 
(see section \ref{subsect:QC_cal}).

In the following subsections, we describe the QCs in each of the above-mentioned categories. 
For any practical use, the definition of the flags are summarized in Tables
\ref{tab:QCflags_def_V500}, \ref{tab:QCflags_def_V1200} and \ref{tab:QCflags_def_COMB}.
Parameters that either are involved in the determination of the QC flags or that can be
useful for independent QC assessment by the user are released as "QCpars" tables
available on the CALIFA ftp site, along with a short description of the parameters.

\input{QCflags_def_V500.tex}
\input{QCflags_def_V1200.tex}
\input{QCflags_def_COMB.tex}

\subsection{Quality of the observing conditions (\textsc{obs})}\label{subsect:QC_obs}

Three quantities contribute to determine the quality of the observing conditions of the CALIFA data: 
the airmass, the sky brightness, and the atmospheric extinction. 
Contrary to DR2, guided by our increased experience,
in DR3 observing conditions never raise a \textsc{bad} flag:
in other words, poor observing conditions alone do not imply bad or unsuitable
data, but just raise a \textsc{warning}.

Note that, as in DR2, we do not consider the seeing in the set of observing condition
parameters to set flags, although it is included in the released "QCpars" tables.
The reason for this is that, as already pointed out in \citet{rgb15},
given the sampling of the fibers on the plane of the sky and the resampling
process, the resolution and depth of the CALIFA cubes are largely insensitive
to the seeing during the observations.

For the airmass, we consider the average and the maximum airmass of the observations
over all contributing pointings (\textsc{obs\_airmass\_mean} and 
\textsc{obs\_airmass\_max}) and its rms (\textsc{obs\_airmass\_rms}). 
For each of these quantities, we defined a warning threshold, which 
is more restrictive in V1200 than in V500 due to the more demanding 
observing conditions for V1200 (see Tables \ref{tab:QCflags_def_V500} and 
\ref{tab:QCflags_def_V1200}). The combined \textsc{flag\_obs\_am} results 
in a \textsc{warning} as long as any of the three quantities exceeds the threshold.

The V-band surface brightness of the sky during the observations (\textsc{skymag})
may limit the depth of the observations and the accuracy of the 
sky subtraction and therefore contributes in defining the quality
of the observing conditions. As for DR2, the quantity \textsc{skymag} is measured in each 
pointing from the sky spectrum obtained from the 36 sky fibers\footnote{See Appendix A.8 
of \citet{Husemann:2013}.}. The mean and the rms of \textsc{skymag} over
all pointings are considered to define the corresponding flags. As for the airmass, stricter 
requirements are applied to V1200 data than to the V500 data. Note also that the thresholds are 
different from those adopted in DR2, as a result of our improved understanding of the impact of
the sky brightness on our data.

The transparency of the sky during each pointing (\textsc{ext}) is 
obtained from the monitored $V$ band extinction at the time of the observation. 
Large extinctions on average, a large maximum extinction or a large rms variation
across the pointings (indicating inhomogeneous observing conditions) set a
\textsc{warning} flag, according to the thresholds reported in Tables 
\ref{tab:QCflags_def_V500} and \ref{tab:QCflags_def_V1200}. 

\subsection{Quality of the instrumental/data reduction performance (\textsc{red})}\label{subsect:QC_red} 

We assess the quality of the instrumental and data reduction performance based
on four different properties as measured on the reduced data {\em before} combining
them into the final datacube: \textsc{straylight}, spectral \textsc{dispersion},
cross dispersion \textsc{cdisp}, and the residuals from the subtraction of bright skylines 
(namely, the 5577 \AA~O$_2$~line in the V500 setup and the 4358 \AA~Hg\textsc{i}~in
the V1200 setup). Moreover, we consider the limiting surface brightness 
corresponding to a 3-$\sigma$ detection per spaxel and spectral 
resolution element measured on the final datacube.
Additionally we check that the final datacube does not present a pathological
fraction of bad pixels, i.e. pixels characterized by large errors (5 times larger than the absolute value of the flux).
Thresholds on this fraction \textsc{red\_frac\_bigerr} are given in order to raise
a \textsc{warning} or a \textsc{bad} \textsc{flag\_red\_errspec} flag.

The straylight is an additive contribution to the raw spectra that must be removed
in the data reduction process. Although the pipeline takes care of the straylight subtraction
\citep[see Appendix A.3 of ][]{Husemann:2013}, we have found that significant residuals 
that affect the final quality of the data are left whenever a frame presents high mean levels of 
straylight (\textsc{meanstraylight}), as well as high maximum values (\textsc{maxstraylight}) 
and large rms (\textsc{rmsstraylight}). Tables \ref{tab:QCflags_def_V500} and \ref{tab:QCflags_def_V1200} 
report the thresholds above which a \textsc{warning} or a \textsc{bad} 
\textsc{flag\_red\_straylight} flag is set, respectively, for the three quantities
in any of the contributing 2D frames (as indicated by the \textsc{\_max} suffix
attached to each quantity). Note that we have modified the thresholds adopted in the DR2
to less strict values, based on the larger statistics now available and our greater experience with the data. 
{ Specifically, while the original thresholds were set based on the distributions of the parameters
and the corresponding percentiles in order to filter out clear outliers, in this release we have
anlayzed the actual correspondence of \textsc{warning} and \textsc{bad} flags to real problems in the cubes,
thus realizing that the requirements on the straylight for an acceptable reduction could be relaxed.}

The light from each fiber is dispersed in the wavelength direction with 
a given spectral \textsc{dispersion} along a trace with
a finite width or cross-dispersion FWHM (\textsc{cdisp}). Significant departures of
these two quantities from the nominal target values raise a \textsc{flag\_red\_disp}
and a \textsc{flag\_red\_cdisp} flag, respectively. This is done by checking
the mean values (\textsc{red\_disp\_mean}, \textsc{red\_cdisp\_mean}), the rms
(\textsc{red\_disp\_rms}, \textsc{red\_cdisp\_rms}), and the maximum values
(\textsc{red\_disp\_max}, \textsc{red\_cdisp\_max}) against the thresholds provided
in Tables \ref{tab:QCflags_def_V500} and \ref{tab:QCflags_def_V1200} 
\citep[see footnote 10 
in][for more details about these quantities]{rgb15}. Again, note that
the thresholds have been modified with respect to DR2 in order to optimize the
effectiveness of the flags. { Specifically, we have relaxed the requirements on the maximum values and rms, which could be strongly affected by a few low-quality spaxels, even if the cube has a generally good quality.}

The accuracy of the sky subtraction is quantified by the minimum and the maximum over all 
pointings of the average (over all fibers) flux residual of a bright skyline within an individual pointing 
(\textsc{red\_res4358\_min} and \textsc{red\_res4358\_max}, and \textsc{red\_res5577\_min} and 
\textsc{red\_res5577\_max} for the V1200 and the V500 setup, respectively). We also consider the maximum 
over all pointings of the rms residuals (over all fibers in an individual pointing), 
\textsc{red\_rmsres4358\_max} and \textsc{red\_rmsres5577\_max}. Large average residuals 
(in absolute value) are indications of systematic bias in the sky subtraction, while large rms
is a symptom of localized failures or noisy data. In these cases, the \textsc{flag\_red\_skylines} 
is set.

The \textsc{flag\_red\_limsb} flag is used to classify the quality of datacubes based on 
the 3-$\sigma$ continuum flux density detection limit per interpolated 1 arcsec$^2$-spaxel 
and spectral resolution element. See Section \ref{sect:depth} for a definition of the 
wavelength range used to derive this quantity. Thresholds are provided in AB-magnitudes
over the spectral window used for the flux integration and have been tuned slightly 
with respect to DR2.

\begin{figure}
\resizebox{\hsize}{!}{\includegraphics{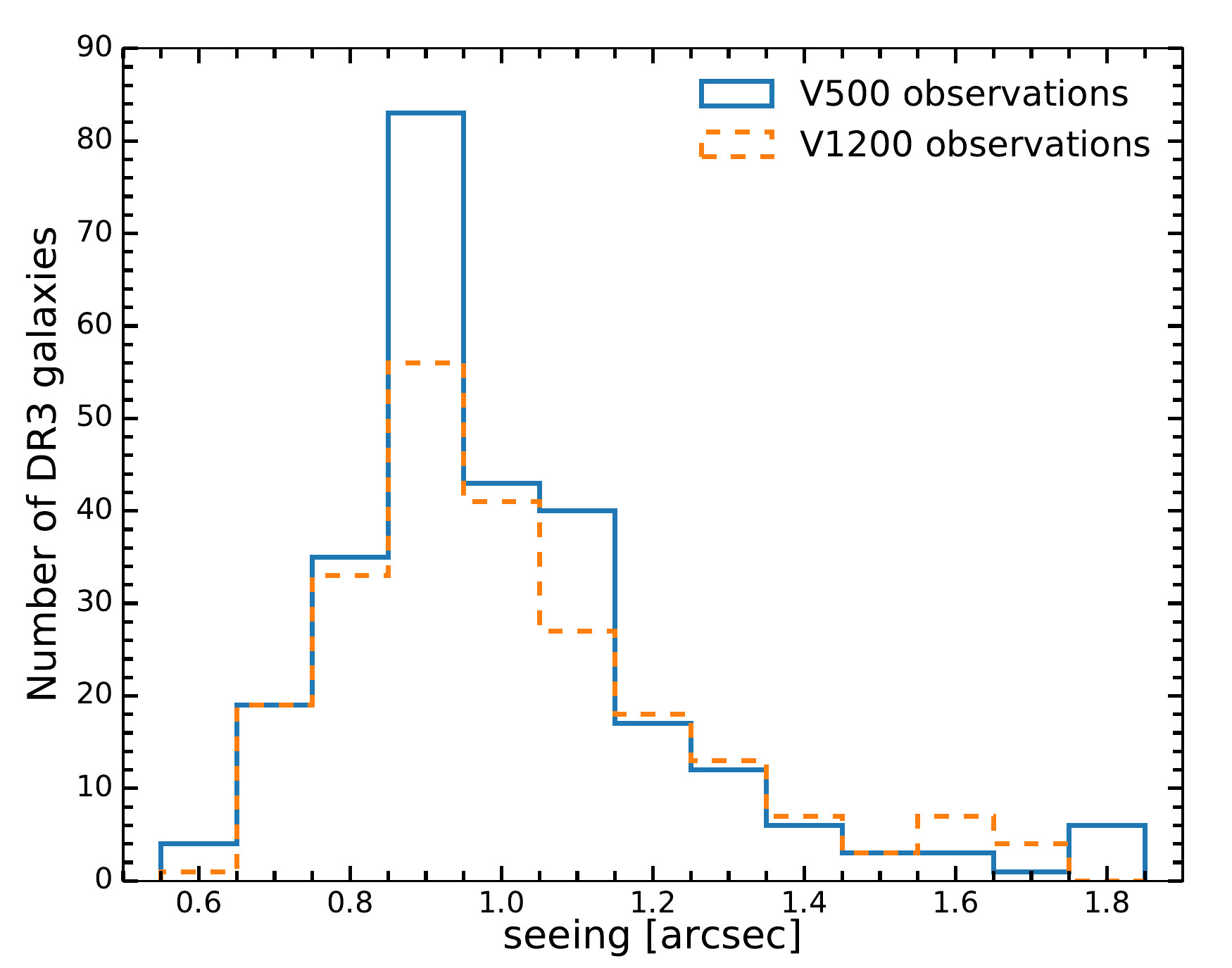}}
\caption{Distribution of the seeing during the CALIFA observations as measured
  by the automatic Differential Image Motion Monitor
  \citep[DIMM,][]{Aceituno:2004}.}
  \label{fig:DR3_seeing}
\end{figure}

\subsection{Quality of the calibrated data products (\textsc{cal})}\label{subsect:QC_cal}

This part of the QC deals with the final calibrated datacubes and, although similar
in many respects, has been significantly modified with respect to DR2.

A series of checks and flags are common to both the V500 and V1200 setups, namely those concerning
the quality of the synthetic image (\textsc{flag\_cal\_imgqual}), of the 
integrated 30\arcsec-aperture spectra (\textsc{flag\_cal\_specqual}), the
wavelength accuracy and stability (\textsc{flag\_cal\_wl}), and, when applicable,
the quality of the "SDSS flat-field" (see below and Section \ref{sect:SDSSflat}) and of the
registration on the plane of sky relative to SDSS imaging \citep[see below and][]{rgb15}.
For V500 only, in addition we perform checks on the spectrophotometric accuracy
that determine the \textsc{flag\_cal\_specphoto} flag. Additional checks are
visually performed on the synthetic images and the 30\arcsec-aperture spectra
for COMBO cubes. Wavelength accuracy and stability (\textsc{flag\_cal\_wl}) are performed on
COMBO cubes as well.

Visual checks on the reconstructed synthetic images in $V$-band (V500 and COMB)
and B band (V1200 and COMB) are encoded in the \textsc{flag\_cal\_imgqual}
flag and take into account the quality of the images in absolute terms and in comparison to
the corresponding $g$-band SDSS image. Reasons to raise a \textsc{warning} or a \textsc{bad} flag
are: holes, artefacts of any kind, irregular background, ghosts, evident noise patterns,
duplicate/offset images of the same sources, significantly elongated PSF.
A \textsc{bad} \textsc{flag\_cal\_imgqual} flag
implies that the datacube is not suitable for any scientific use and occurs whenever
multiple images are spotted, indicating a serious problem in the (relative) registration
of the pointings, or when a substantial fraction of the image is affected by anomalous
background subtraction. Noise patterns or background problems over a limited region are usually
flagged as \textsc{warning}.

Visual checks on the 30\arcsec-aperture spectra are meant to flag anomalies in the overall spectral shape,
such as bumps, drops, anomalously low SNR (possibly in limited spectral ranges), and are encoded into the 
\textsc{flag\_cal\_specqual} flag. A \textsc{bad} classification in this flag implies that
the cube is not useful for any science and therefore is not released.

Whenever possible, pipeline V2.2 renormalizes the spectra in each spaxel to match the
photometric fluxes derived from the co-registered SDSS images, by applying the SDSSflat correction. 
The map of the SDSS flat-fielding function is visually inspected, as well as
are the synthetic images and the 30\arcsec-aperture integrated spectra 
before and after applying the SDSSflat. Since the reduced data cubes are already flux-calibrated
before applying the SDSSflat, the correction applied in this step must be small, a few percent
at most. Large correction factors and/or strong spatial variations, possibly due to bad CALIFA versus SDSS 
co-registration (spatial offsets or badly matched spatial resolution) and resulting
in large differences between the integrated 
spectra before and after the SDSSflat correction, are initially marked with a \textsc{bad} 
\textsc{flag\_cal\_flatsdss}. We have then re-checked all such cases and investigated
if the problem is due to an independent failure in the observations/data reduction or if it is due to
the SDSSflat itself: in the latter case the datacube is re-reduced without the SDSSflat
correction and the \textsc{flag\_cal\_flatsdss} is assigned undefined ($-1$). There are cases
where the problem is judged as not amendable and therefore a \textsc{bad} 
\textsc{flag\_cal\_flatsdss} is retained. If the problem is flagged as \textsc{warning} 
the SDSSflat correction is retained.

Pipeline V2.2 by default attempts to register the different pointings relative to the SDSS
imaging before combining them. The outcome of the process is checked in the QC protocol by
visually inspecting i) the footprints of the real CALIFA fibers and of the simulated fibers on the
SDSS images based on the matched astrometric solution and ii) the $\chi^2$ surfaces that describe the
goodness of the match between SDSS and CALIFA as a function of the 2D spatial offsets.
These checks result in the \textsc{flag\_cal\_registration} flag. If a \textsc{bad}
condition occurs (i.e. obvious mismatch in the pointings relative to SDSS), the datacube is
inspected again and re-reduced with the registration based
on relative telescope offsets as in the V1.3 pipeline. Section \ref{sect:pipeline} and 
\citet{rgb15} contain more information about
the main differences between the two methods. Should this re-reduction produce acceptable
results in terms of image and spectral quality, the \textsc{flag\_cal\_registration} is
assigned undefined ($-1$) and the datacube is released, otherwise the \textsc{flag\_cal\_imgqual} and/or
\textsc{flag\_cal\_specqual} flags are assigned \textsc{bad} and the datacube is rejected. 

The QC protocol foresees a-posteriori flux calibration checks for the V500 setup only,
which determine the flag \textsc{flag\_cal\_specphoto}. A first quantitative
check relies on the visual inspection of the 30\arcsec-integrated spectra, whereby the $g$- and $r$-band 
magnitudes derived from SDSS images integrated over the same area are transformed to flux densities and
overplotted on the CALIFA spectrum: significant offsets between the SDSS points and the spectrum
raise the flag. The flag is also raised in case of visual checks revealing anomalous spectral shapes (bumps,
drops etc.). Finally and more quantitatively the flux ratios in $g$- and $r$-band of the different pointings
relative to SDSS are considered: deviations in the mean and/or the rms of the flux ratios over the different
pointings by more than given thresholds raise the corresponding flag. The flag \textsc{flag\_cal\_specphoto} 
eventually reports the worst classification extracted from all these checks.

In order to check the stability of the wavelength calibration over the full spectral range we 
performed the same measurements as in DR1 and DR2, as described in Section 5.3 of \citet{Husemann:2013} 
\citep[see also][]{rgb15}: for each 
galaxy and setup, the spectra within 5\arcsec\ of the center of the galaxy are integrated and the 
systemic velocity is estimated first for the full spectrum and then for 3 (4) independent spectral 
ranges in the V1200 (V500) setup. The rms of these values with respect to the systemic velocity from the full 
spectrum (\textsc{cal\_rmsvelmean}) is an estimate of the stability of the wavelength calibration 
across the wavelength range and is used to set the corresponding quality flag \textsc{flag\_cal\_wl}.

\subsection{Overall quality assessment}
\label{sect:QC_select}
The flags described in the previous sections allow any potential user to select samples that are
most suitable for her/his science goals, using ad hoc selection criteria.
However, we identify a set of key flags for which a \textsc{bad} classification implies
unusable data in any respect: cubes with a \textsc{bad} in either \textsc{flag\_cal\_imgqual},
\textsc{flag\_cal\_specqual}, \textsc{flag\_cal\_wl} or \textsc{flag\_red\_disp} are therefore
excluded from DR3. The sample of galaxies with a released datacube in either one or both of 
the V500 and V1200 setups satisfying these criteria has been defined as the CALIFA DR3 
in Section \ref{sect:DR3_sample}. 

If one wishes to be more strict, one could restrict all flags to \textsc{warning} at most (value $\leq 1$). Such a selection produces a sample we have called the high-quality sample (HQ sample) in Section \label{s:hqsamp}, containing 332 galaxies. An even more stringent restriction would additionally require perfect quality ($0$ value) in the key flags (\textsc{flag\_cal\_imgqual}, \textsc{flag\_cal\_specqual}, \textsc{flag\_cal\_wl}, \textsc{flag\_red\_disp}). Such a selection would produce a sample of 124 galaxies with the highest quality data. That sample will be limited in size and therefore of less scientific use, but could still be used as a reference sample for making sure that no data imperfections affect scientific conclusions derived from a specific method or paper.

\begin{figure}
\resizebox{\hsize}{!}{\includegraphics{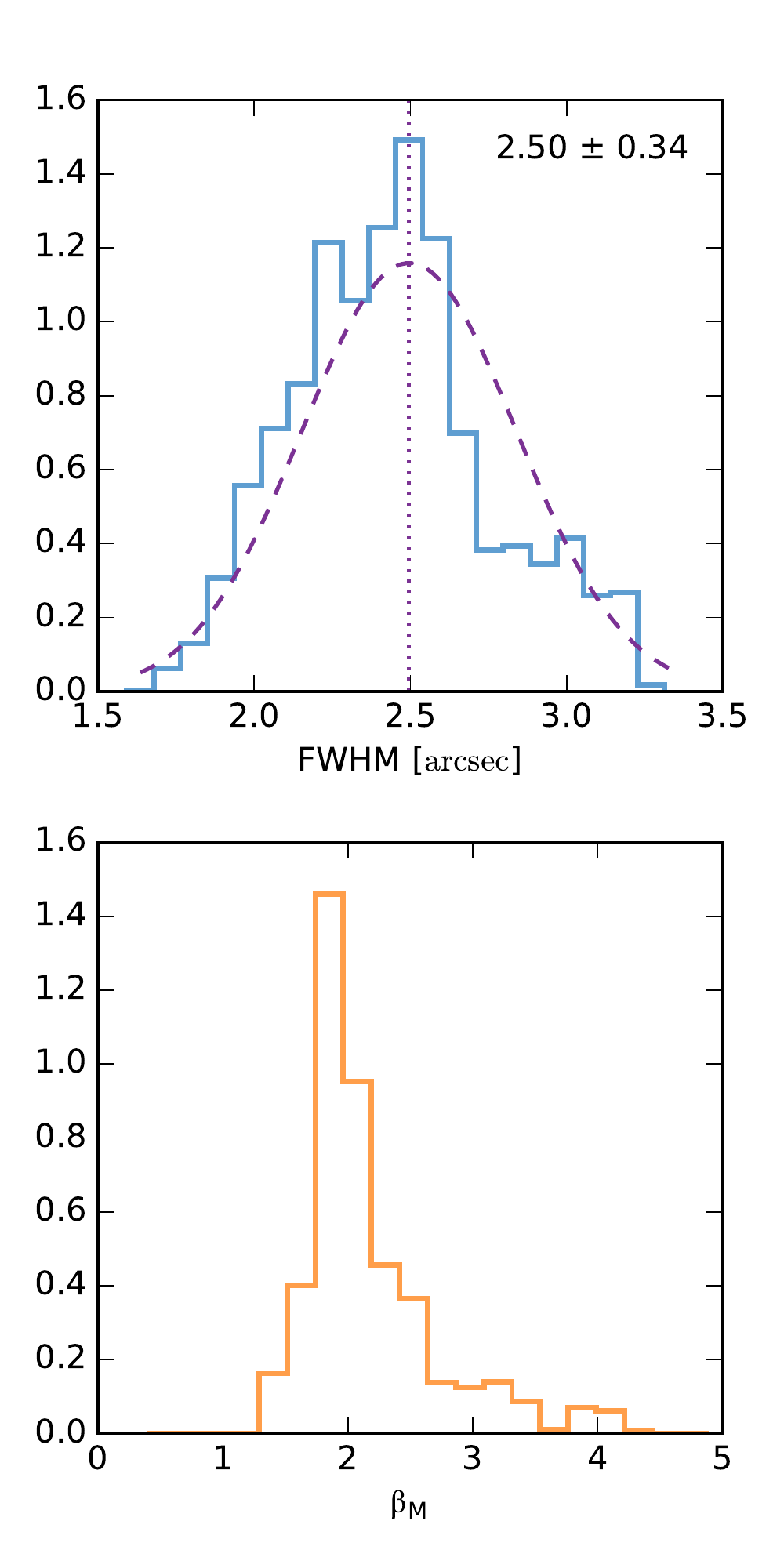}}
\caption{Normalized distributions of the integrated area of the PSF FWHM \emph{(top panel)} and $\beta_M$ \emph{(bottom panel)} parameters of an elliptical  2D Moffat profile fitted to 63 calibration stars, weighted by the likelihood of the fit. In the top panel the normal probability density function is marked with a dashed line and the dotted line indicates the mean value of the distribution.}
  \label{fig:DR3_PSF}
\end{figure}

\subsection{Seeing and spatial resolution}\label{sect:spatial_resolution}

The average atmospheric seeing conditions during the CALIFA observations were derived from the measurements 
acquired by the Differential Image Motion Monitor \citep[DIMM,][]{Aceituno:2004}, which operates fully automatically 
at the Calar Alto observatory during the night. The DIMM has different operational constraints from the 3.5m 
telescope (humidity lower than 80\% and wind speed less than $12\,\mathrm{m}\,\mathrm{s}^{-1}$). Seeing 
information is thus not available for every CALIFA observation, but the overall seeing distribution is not 
expected to be very different \citep[see footnote 12 of][]{rgb15}. 

Figure~\ref{fig:DR3_seeing} shows the DIMM seeing distribution for the DR3 sample, which has a median value of 
$1\farcs0$ FWHM, and therefore atmospheric seeing is not a limiting factor in the spatial resolution of the CALIFA 
cubes. Therefore, the final spatial resolution of the CALIFA data is mainly set by fiber size and the dithering and 
interpolation scheme.

We used the following approach to measure the PSF in the datacubes. Since January 2012 standard stars were observed 
using the same dithering pattern adopted for the science observations for both setups. Only a fraction of the nights 
had weather conditions good enough to acquire a calibration star using this scheme, yielding a total of 182 stars 
observed using the dithering scheme. We reduced these data using the same procedure described before for the science 
objects. The PSF can be measured very precisely because the calibration stars have a very high S/N. 
We took images based on slices of width 400 $\AA$ in wavelength from the datacubes for each of these stars. 
For each of these images, we fitted a 2D Moffat profile (see Equation \ref{eq:Moffat}) 
using the software IMFIT \citep{Erwin:2015}. Figure \ref{fig:DR3_PSF} 
shows the normalized distributions of FWHM and $\beta_M$ parameters of the Moffat profile, weighted by the 
likelihood of the fit, for all wavelengths and all stars. Counts are normalized to form a probability density 
so the integral of the histogram is 1. The fits do not show any significant wavelength dependence 
in any of those parameters. We obtained a mean value and 1-$\sigma$ scatter of the FWHM as 2.50 $\pm$ 0.34 arcsec. 
The distribution of $\beta_M$ is asymmetric, so a better estimate of its value is the weighted median, which gives 
$\beta_M$ = 2.15. The ellipticity ($1 - b/a$, with $a$ and $b$ being the semimajor and semiminor axes, respectively) is 
also measured, with mean value and 1-$\sigma$ scatter of 0.08 $\pm$ 0.06. Given the uncertainties, this value 
means the PSF can be considered effectively axisymmetric. The uncertainties in these measurements correspond to 
1-$\sigma$ of the distributions. Note that the distribution is broader than that reported for DR2 
\citep{rgb15}, because in that release we discarded galaxies 
observed under observing conditions with high seeing.

\begin{figure*}
 \includegraphics[width=0.5\textwidth]{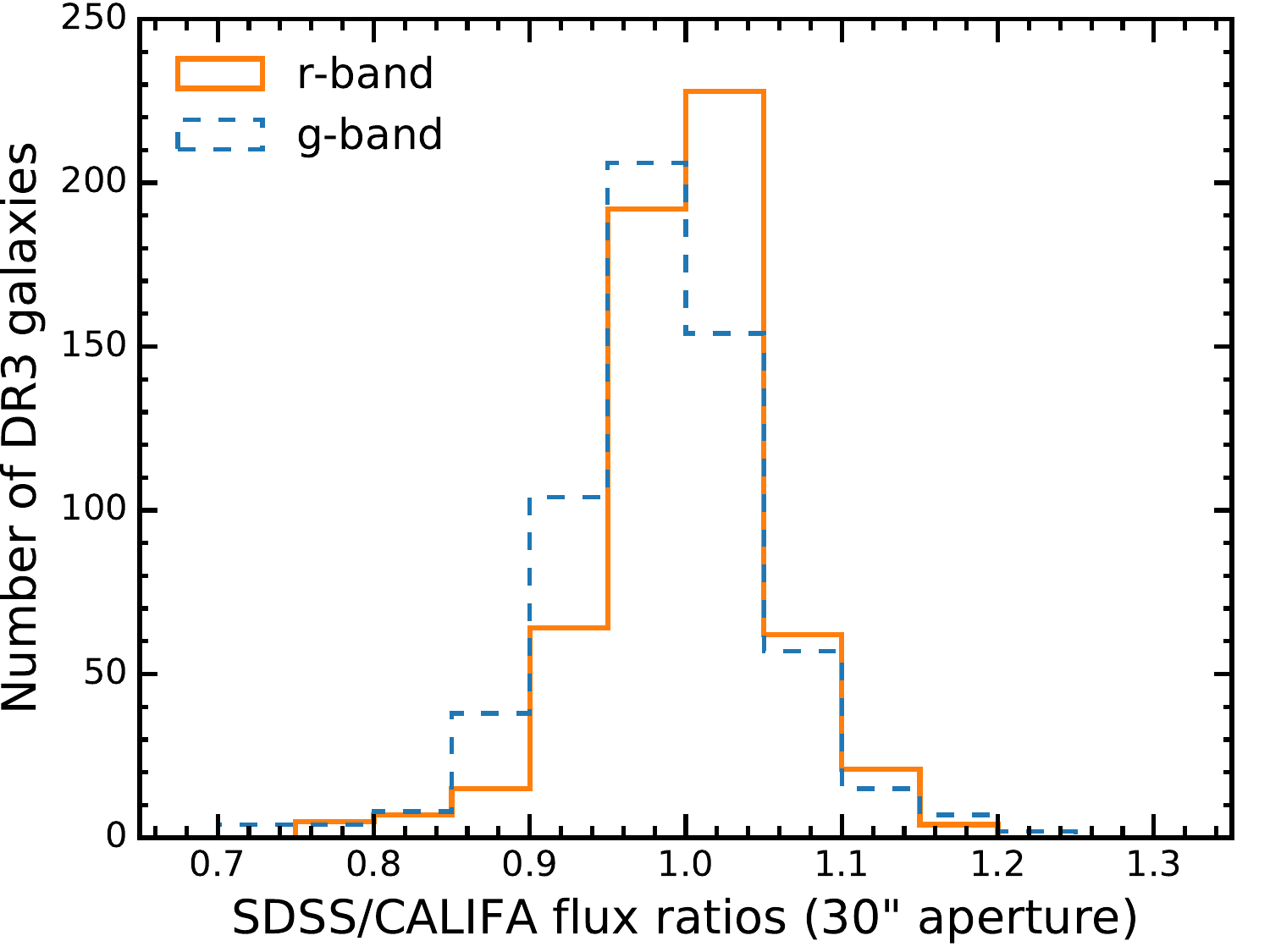}
 \includegraphics[width=0.5\textwidth]{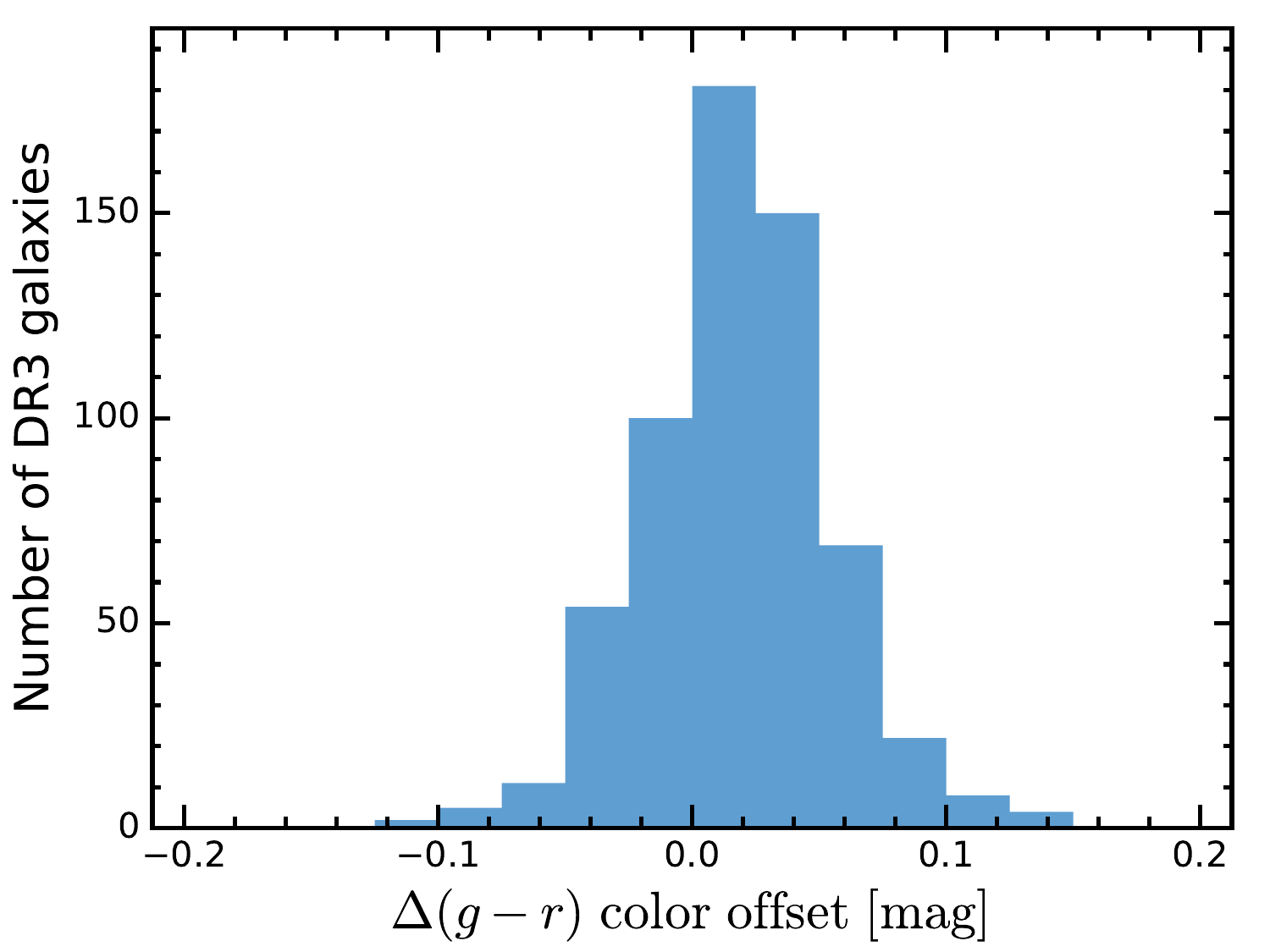}

 \caption{\emph{Left panel:} distribution of the 30\arcsec\ aperture
   photometry scale factor between the SDSS DR7 images and recalibrated
   CALIFA data. We compare the photometry only for the $g$- and $r$-bands,
   which are both entirely covered by the V500 wavelength range. \emph{Right
     panel:} distribution of the corresponding color offset between the SDSS
   DR7 images and the synthetic CALIFA broadband images.}
  \label{fig:DR3_scale_check}
\end{figure*}


\subsection{Spectrophotometric accuracy}\label{sect:specphot_cal} 

As described in Section \ref{sect:reducview} the registration scheme of the pipeline rescales the datacubes 
to the absolute flux level of the SDSS DR7 broad-band photometry, using the $r$-band image for the V500 setup 
and the $g$-band image for the V1200 setup. On the other hand, for the COMBO cubes the V1200 data are finally matched 
to the V500 data. These procedures, together with the recalibrated sensitivity curve (see Section 
\ref{sect:pipeline}), and the updated calibration frames (master skyflats, master bias...) 
improves the spectrophotometric calibration of DR3 relative to DR1 and DR2. 

This is clearly shown in Figure \ref{fig:DR3_scale_check}. As part of the CALIFA pipeline V2.2, a 30\arcsec\ 
diameter photometric aperture in $r$ and $g$ is measured both in the SDSS DR7 images and in the equivalent 
synthetic CALIFA broadband images. The mean SDSS/CALIFA $g$- and $r$-band ratios in DR3 and their scatter are 
0.99 $\pm$ 0.09 and 1.00 $\pm$ 0.08, respectively. In the \emph{right panel} of Figure \ref{fig:DR3_scale_check} 
the distribution in $\Delta(g-r)$ color difference between the SDSS and CALIFA data shows that the 
spectrophotometric accuracy over the wavelength range is better than 4\%, with a median value of 0.02 $\pm$ 0.04.

\begin{figure*}
\resizebox{\hsize}{!}{\includegraphics{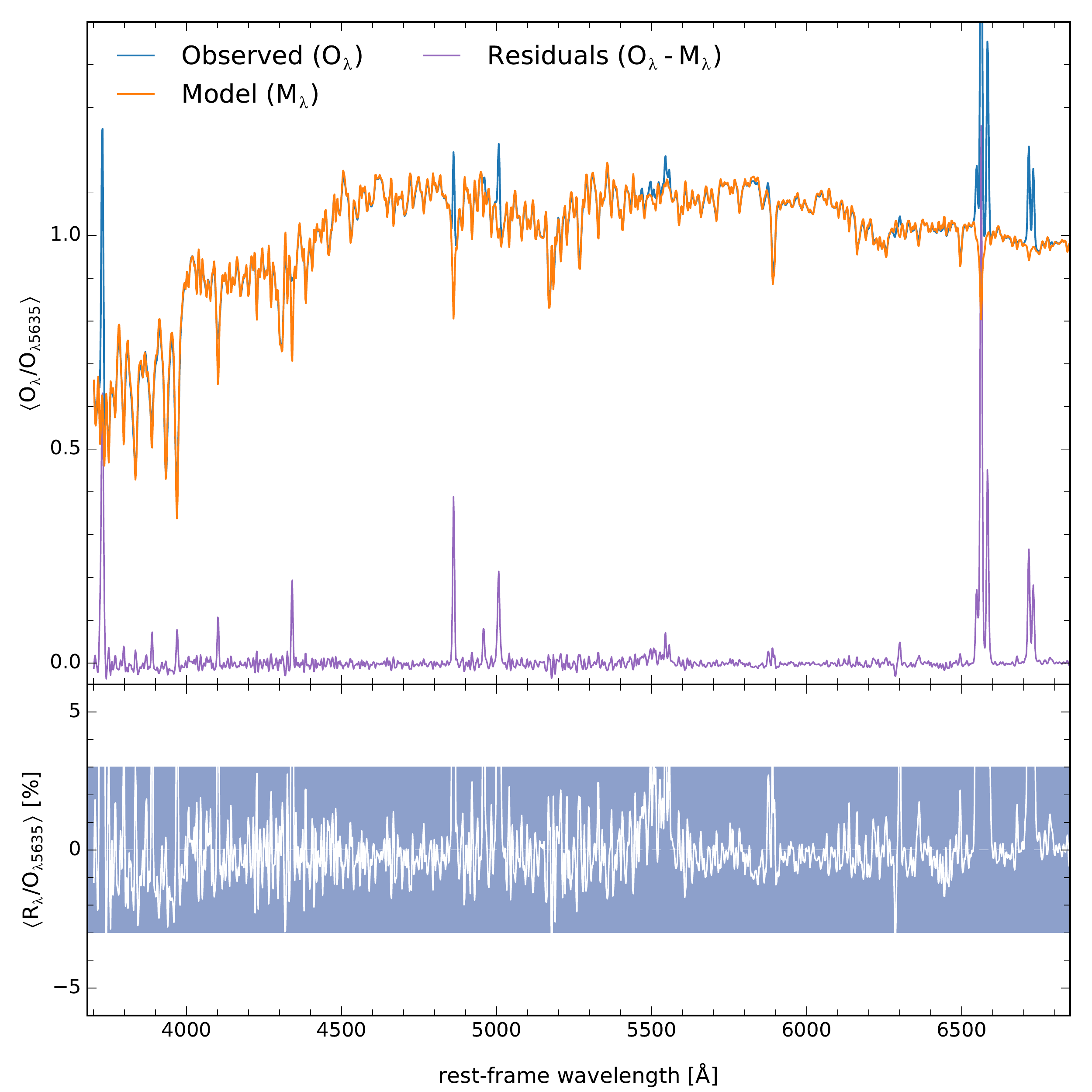}}
 \caption{Statistics of the spectral residuals  {\em Top panel:}  The mean normalized spectrum of 251313 bins from 446 galaxies. The mean {\sc Pipe3D} fit is overplotted   in orange, while the mean residual is plotted at the bottom of the panel (purple).     {\em Bottom panel:} Zoom of the residual spectrum. The shaded rectangle encompasses the $\pm$ 3\% area.}
  \label{fig:DR3_residuals}
\end{figure*}

We use spectral fitting methods to make an independent estimate of the spectrophotometric accuracy, following
\citet{Husemann:2013} and \citet{rgb15}. We repeat a similar experiment for the DR3 datacubes, but in this 
case we use the results from the fitting performed by {\sc Pipe3D} \citep{sanchez2016b}. Results are shown in  
Figure \ref{fig:DR3_residuals}. The top panel shows in blue the mean spectrum of 251313 spatial bins of 
446 galaxies included in the DR3 COMBO distribution with $S/N>$15 in the continuum at $\sim 5635$ \AA\ 
and good quality spectral fitting. The average is taken after normalizing each spectrum  by its 
median flux in the $5635 \pm 45$ \AA\ window. The mean synthetic spectrum (overplotted orange line) as well 
as the mean residual (at the bottom of the upper panel, purple line) are also plotted. The bottom panel zooms 
in on the residual spectrum.

The layout of Figure \ref{fig:DR3_residuals} is similar to Figure 13 in \citet{CidFernandes:2014} or 
Figure 18 in \citet{rgb15}, to which it  should be compared. Focusing on the middle panel, one sees that from 
$\sim 5000$ \AA\ to the red the residuals are very similar, including the humps around 5800 \AA\ associated 
with the imperfect removal of telluric features. Toward the blue however, the reduction pipeline leads to 
smaller residuals than that of version V1.3c, with characteristics very similar to those of version V1.5.


In addition to the previous test, we also performed an independent estimation of the accuracy of the spectrophotometric 
calibration by comparing the flux-calibrated spectra of the stars observed using the three dithering procedures 
(Section \ref{sect:spatial_resolution}) with their published spectra. Since all those stars are spectrophotometric 
standards, they have high-S/N and good quality published spectra.

\begin{figure}
\resizebox{\hsize}{!}{\includegraphics{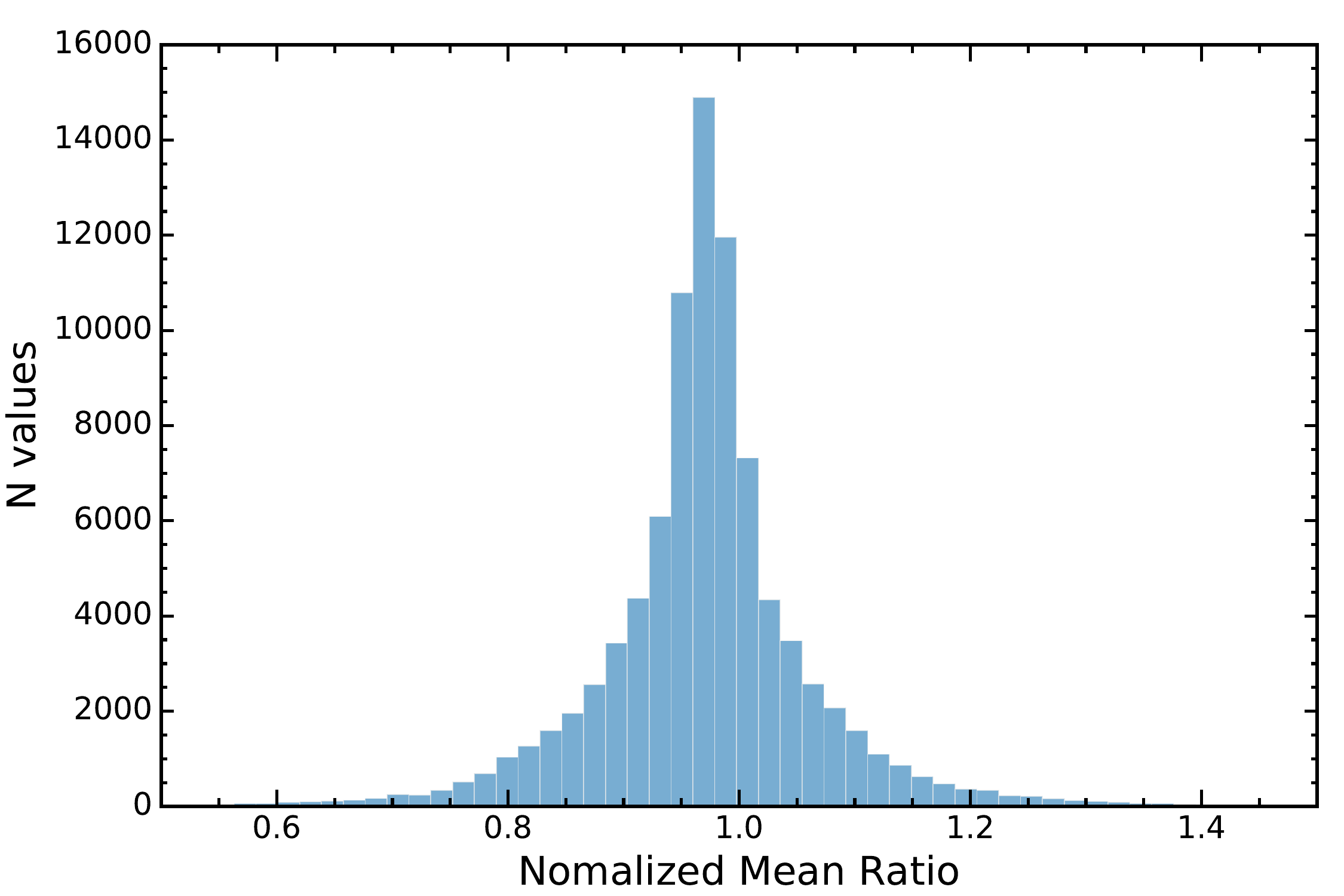}}
 \caption{Histogram of the differences between flux intensities for the published spectra for the spectrophotometric calibration stars described in Section \ref{sect:spatial_resolution} and the corresponding ones derived by CALIFA. }
  \label{fig:hist_cal}
\end{figure}

\begin{figure}
\resizebox{\hsize}{!}{\includegraphics{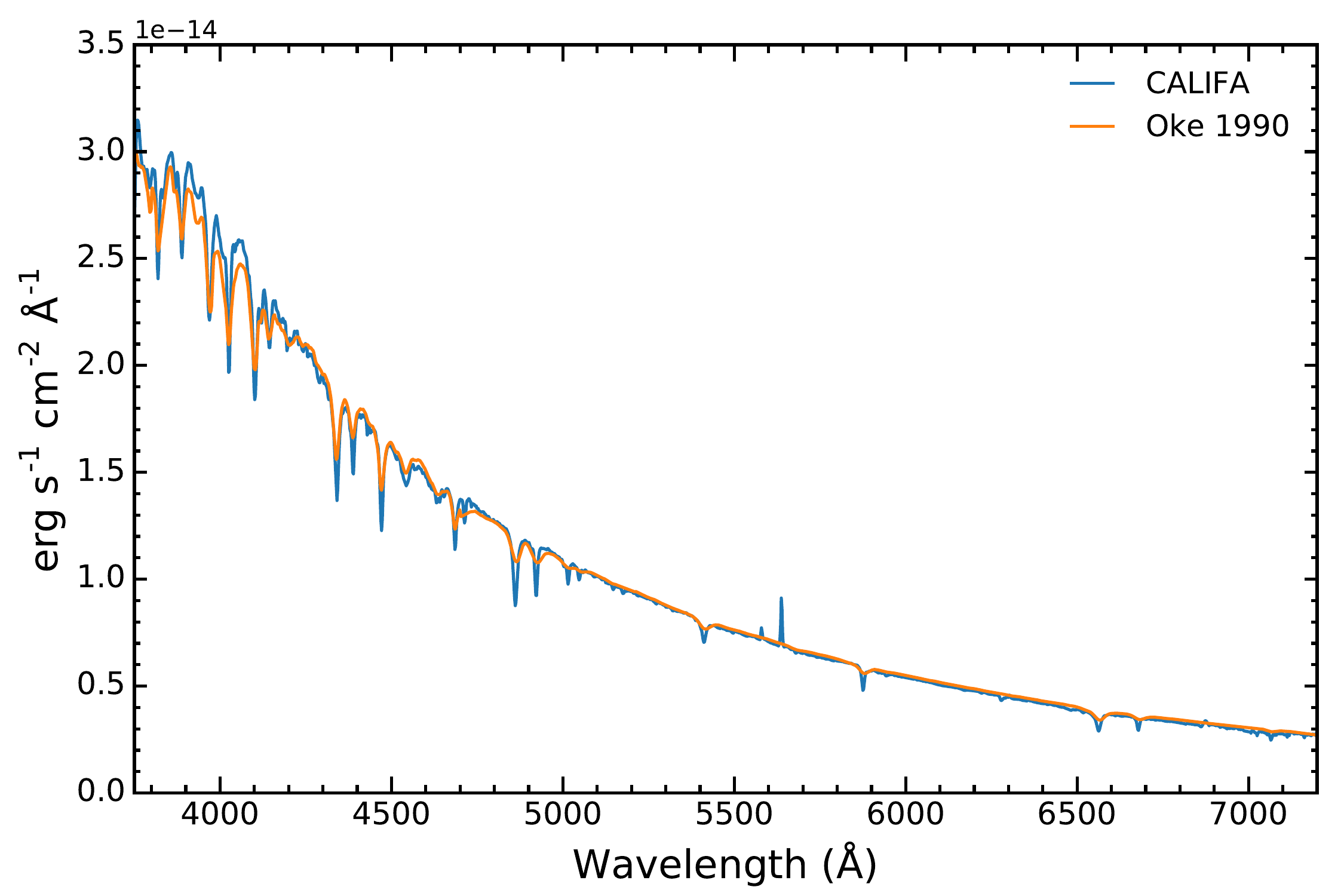}}
 \caption{Comparison between the published spectrum for the spectrophotometric standard star Hz44 and the average of the 15 spectra derived by CALIFA for the same star.}
  \label{fig:spec_Hz44}
\end{figure}

We performed the same spectrophotometric calibration for these stars than the one applied to the science cubes, 
using the same sensitivity curve and atmospheric extinction (Section \ref{sect:pipeline}). Then we extracted a 
30$\arcsec$ aperture spectrum over the DAR-corrected datacube corresponding to each calibration star. Finally we 
compared those spectra with the published ones, deriving a ratio of 0.99$\pm$0.10, as can be seen in 
Figure \ref{fig:hist_cal}. Thus, the absolute spectrophotometric accuracy is around $\sim$10\%, a result that 
was anticipated by \citet{rgb15}. This was the reason why we decided not to use the calibration stars to derive 
the sensitivity curve. Once corrected for the absolute spectrophotometric offset, the average spectra derived 
for each calibration star agree with the published ones within $\sim$3.4\% from blue to red. Figure 
\ref{fig:spec_Hz44} shows the comparison between the derived spectrum of the spectrophotometric standard 
star Hz44 and the published one \citep{Oke90}, showing a high degree of agreement.

\subsection{Limiting sensitivity and signal-to-noise}\label{sect:depth}

\begin{figure}
 \resizebox{\hsize}{!}{\includegraphics{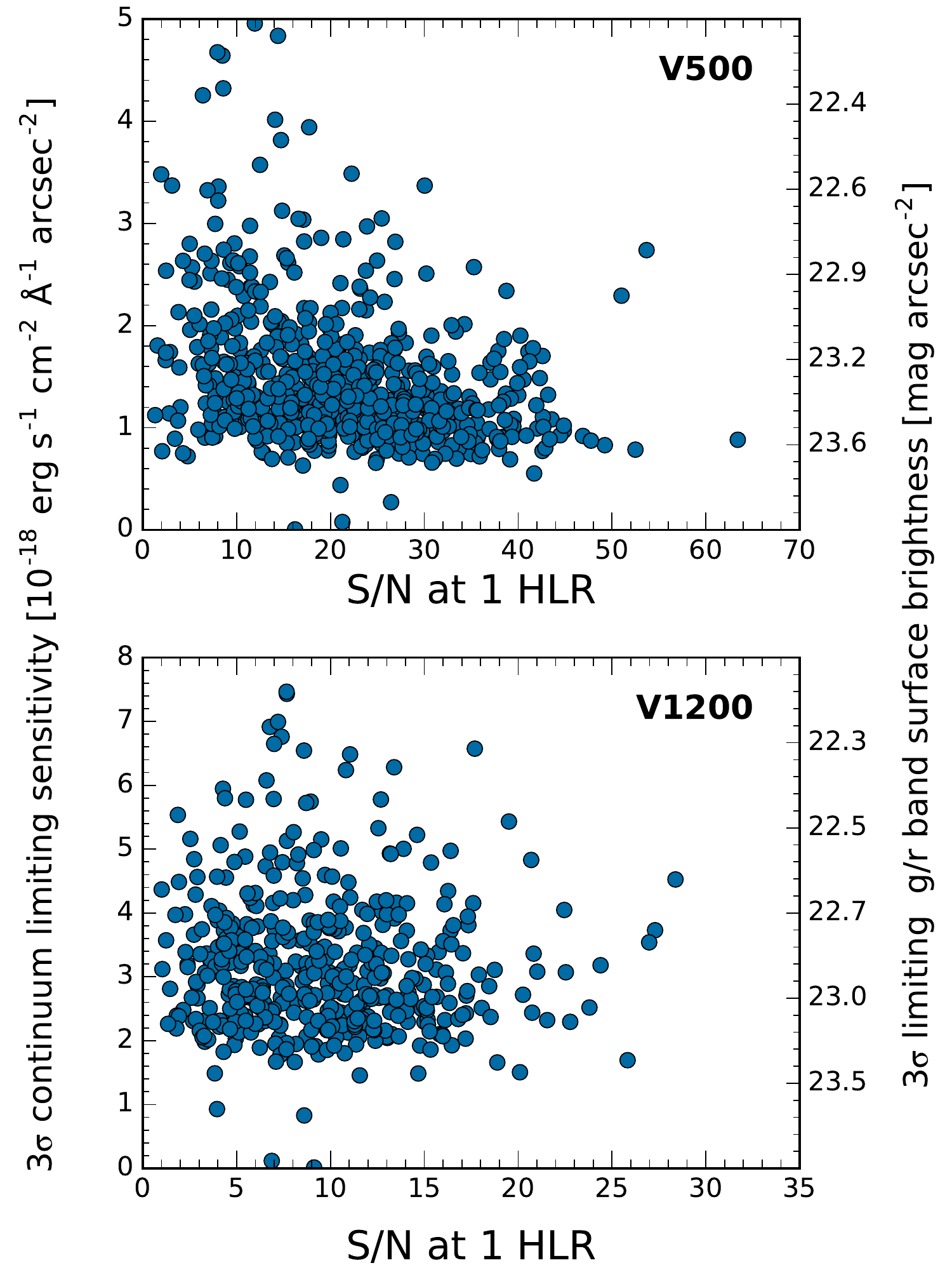}}
 \caption{Limiting 3-$\sigma$ continuum sensitivity per spaxel and spectral resolution element as a 	function of the average continuum S/N at the half-light radius (HLR). The corresponding broadband 
	surface brightness limits in $r$ (V500) and $g$ (V1200) are indicated on the right $y$-axis. The 	limiting continuum sensitivity and the S/N were computed from the median signal and noise in the 	wavelength region 4480--4520 \AA\ and 5590--5680 \AA\ for the V1200 and V500 data, respectively.}
 \label{fig:DR3_depth}
\end{figure}

To assess the depth of the data, we estimate the 3-$\sigma$ continuum flux density detection 
limit per interpolated $1\,\mathrm{arcsec}^{2}$-spaxel and spectral resolution element for the 
faintest regions. Figure \ref{fig:DR3_depth} shows the limiting continuum sensitivity of the 
spectrophotometrically recalibrated CALIFA cubes. The depth is plotted against the average S/N 
per 1 arcsec$^{2}$ and spectral resolution element within an elliptical annulus of $\pm1\arcsec$ 
around the galaxies' $r$-band half-light semimajor axis (HLR), with position angle (PA) and radius values taken from 
\citet{Walcher:2014} when available or directly from the datacube. A narrow wavelength window at 
4480--4520 \AA\ for the V1200 and at 5590--5680 \AA\ for the V500 
is used to estimate both values. Specifically, the signal (also used for the surface brightness limit) 
is computed as the median value in the defined wavelength intervals, while the noise is the detrended 
standard deviation in the same windows. These small windows are nearly 
free of stellar absorption features or emission lines. The 3-$\sigma$ continuum flux density detection 
limit per spaxel and spectral resolution element\footnote{Note that this is a continuum flux density. 
See Note 5 of \citet{Husemann:2013}.} for the V1200 data 
($I_{3\sigma}=3.0\times10^{-18}\,\mathrm{erg}\,\mathrm{s}^{-1}\,\mathrm{cm}^{-2}\,\mathrm{\AA}^{-1}\,\mathrm{arcsec}^{-2}$
in the median at 4500\AA) is a factor of $\sim$2-3 brighter than for the V500 data
($I_{3\sigma}=1.3\times10^{-18}\,\mathrm{erg}\,\mathrm{s}^{-1}\,\mathrm{cm}^{-2}\,\mathrm{\AA}^{-1}\,\mathrm{arcsec}^{-2}$
in the median at 5635\AA) mainly because of the difference in spectral resolution. 
These continuum sensitivities can be transformed into equivalent limiting broadband surface brightnesses 
of $23.0\,\mathrm{mag}\,\mathrm{arcsec}^{-2}$ in the $g$-band for the V1200 data and
$23.4\,\mathrm{mag}\,\mathrm{arcsec}^{-2}$ in the $r$-band for the V500 data.
The variance of the sky brightness on each night might be one of the main factors causing the difference in 
the limiting continuum sensitivity. Dust attenuation, transparency of the night, and other atmospheric 
conditions might also affect the depth achievable at fixed exposure times. 

The limiting sensitivity is also a measure of the noise due to observing conditions and thus it correlates 
mildly with the S/N at one HLR. The mean S/N in the continuum per 1 arcsec$^{2}$ and spectral resolution 
element at the HLR along the semimajor axis for all objects is $\sim$9.4 for the V1200 setup, while it 
is $\sim$21.2 for the V500 setup. Thus, we achieve a S/N$\gtrapprox$10 at 1 HLR for a significant number 
of the objects { for the V500 setup ($\sim$85\%) and even for the V1200 setup ($\sim$40\%)}. 


\section{Access to the CALIFA DR3 data}\label{sect:DR2_access}

\subsection{The CALIFA DR3 search and retrieval tool}

The public data are distributed through the CALIFA DR3 web page (\url{http://califa.caha.es/DR3}). A simple web 
form interface, already in use for DR1 and DR2, allows the user to select data of a particular 
target galaxy, or a sub-sample of objects within some constraints on observing conditions or galaxy properties. 
Among the selection parameters, we include the instrument setup, galaxy name and coordinates, redshift, $g$-band 
magnitudes, Hubble type, bar strength and whether or not it is a clearly merging system. 

If any CALIFA data sets are available given the search parameters, they are listed in the search results and can be selected for download. The download process requests a target directory on the local 
machine to store the data, after the downloading option is selected. The CALIFA data are delivered as 
fully reduced datacubes in FITS format separately for each of the three configurations: V500, V1200 and 
COMBO. Each DR3 datacube is uniquely identified by its file name, 
\texttt{$GALNAME$.V1200.rscube.fits.gz}, \texttt{$GALNAME$.V500.rscube.fits.gz} and 
\texttt{$GALNAME$.COMB.rscube.fits.gz} for the V1200, V500 and COMBO configurations respectively, 
where $GALNAME$ is the CALIFA ID number listed in the electronically distributed tables.

All the QC tables discussed throughout this article are also distributed in CSV and FITS-table formats 
on the DR3 webpage. In addition, we distribute the tables discussed in \citet{Walcher:2014} and in 
Section \ref{sect:DR3_sample} regarding the characterization of the Main and Extension Samples, using similar 
formats. These tables could be useful for further science exploitation of the datacubes.

\subsection{Virtual Observatory services}

Just like the previous data releases, the CALIFA DR3 also interfaces with Virtual 
Observatory (VO) facilities.  At release time or shortly
thereafter, the datacubes will be made available through the Obscore data model,
and as database tables of voxels in the TAP service
\texttt{ivo://org.gavo.dc/tap}\footnote{accessible at
\url{http://dc.g-vo.org/tap}}. The service also contains tables of
objects and QC parameters.  These and further services
can also be found by searching for CALIFA DR3 with any
Registry client.


\section{Summary}

In this article we have presented the main characteristics of the third public data release, DR3, of the Calar 
Alto Legacy Integral Field Area (CALIFA) survey. DR3 comprises 667 galaxies (1576 datacubes) 
containing information from more than 1.5 million independent spectra, covering a wide range of masses, 
morphological types, and colors. The released datacubes correspond to two different sample of galaxies: 
i) the Main Sample, a randomly selected sub-sample of the CALIFA mother sample comprising 529 galaxies 
and representative of galaxies in the local universe, and ii) an Extension Sample comprising a 
heterogeneous collection of galaxies observed with the same setup, that add objects that are rare in the 
overall galaxy population and therefore not numerous or absent in the mother sample. The CALIFA DR3 provides 
science-grade and quality-checked integral-field spectroscopic data  
to the community at \url{http://califa.caha.es/DR3}.

We reduced the data using a new version of the pipeline (V2.2), which improves slightly the quality of the data 
in terms of: i) reliability of the spatial registration process, ii) the homogeneity in the data reduction, and 
iii) the quality of the image reconstruction. We described in detail the main quality parameters analysed in 
the validation process, which are provided to users with complete tables to select the most suitable objects 
for their science cases. 

Compared with other ongoing major surveys, CALIFA offers a similar projected spatial resolution. The PSF of 
the datacubes has a mean value of $\sim$ 2.5\arcsec\ (Section \ref{sect:spatial_resolution}), similar to the 
one reported SAMI \citep{Sharp:2014} and MaNGA (Law et al. in prep.). However, CALIFA galaxies are observed 
at lower redshift and with a physically larger IFU, thus providing better spatial coverage and resolution, 
as well as better overall S/N. CALIFA thus offers a highly competitive compromise for analysing the 
spatially resolved structures in galaxies. The penalty for this wider coverage is the lower number of 
galaxies observed (6 times lower than what is projected for SAMI and 15 times lower than the goals of 
MaNGA), and a lower spectral resolution over the full wavelength range.

While the CALIFA data distributed in this final DR have already been used for a variety of science applications, 
the potential for new scientific exploitation of the data is still very large. As CALIFA has been conceived 
as a legacy survey, we sincerely hope that the data will be useful to the community in years to come. 

\begin{acknowledgements}

CALIFA is the first legacy survey being performed at Calar Alto. The CALIFA collaboration would like to thank 
the IAA-CSIC and MPIA-MPG as major partners of the observatory, and CAHA itself, for the unique access
to telescope time and support in manpower and infrastructures. The CALIFA collaboration thanks also the CAHA 
staff for the dedication to this project. 

We thank the anonymous referee for his/her help in improving this article.

SFS thanks the director of CEFCA, M. Moles, for his sincere support to
this project. 

SFS thanks the CONACYT-125180 and DGAPA-IA100815 projects for
providing him support in this study.
RGB, RGD, and EP are supported by grants  AYA2014-57490-P and JA-FQM-2828.
SZ is supported by the EU Marie Curie Integration Grant ``SteMaGE'' Nr. PCIG12-GA-2012-326466 
(Call Identifier: FP7-PEOPLE-2012 CIG).
J.~F-B. from grant AYA2013-48226-C3-1-P from the Spanish Ministry of Economy and Competitiveness (MINECO), as well as from the FP7 Marie Curie Actions of the European Commission, via the Initial Training Network DAGAL under REA grant agreement number 289313
B.G-L- acknowledges financial support by the Spanish MINECO under grants AYA2013-41656-P and AYA2015-68217-P
Support for L.G. is provided by the Ministry of Economy, Development, and Tourism's Millennium Science Initiative 
through grant IC12009, awarded to The Millennium Institute of Astrophysics, MAS. L.G. also acknowledges support 
by CONICYT through FONDECYT grant 3140566.
and AYA2013-42227-P from the Spanish Ministerio de Ciencia e Innovación and TIC 114 and PO08-TIC-3531 from Junta de Andalucía.
AG acknowledges support from the FP7/2007-2013 under grant agreement n. 267251 (AstroFIt).
RAM was funded by the Spanish programme of International Campus of Excellence Moncloa (CEI).
JMA acknowledges support from the European Research Council Starting Grant (SEDmorph; P.I. V. Wild).
IM and AdO acknowledge the support by the projects AYA2010-15196 from the Spanish Ministerio de Ciencia e 
Innovaci\'on and TIC 114 and PO08-TIC-3531 from Junta de Andaluc\'ia.
AMI acknowledges support from Agence Nationale de la Recherche through the STILISM project (ANR-12-BS05-0016-02).
MM acknowledges financial support from AYA2010-21887-C04-02 from the Ministerio de Economía y Competitividad.
PSB acknowledges support from the Ram\'on y Cajal program, grant ATA2010-21322-C03-02 from the Spanish 
Ministry of Economy and Competitiveness (MINECO).
CJW acknowledges support through the Marie Curie Career Integration Grant 303912.
VW acknowledges support from the European Research Council Starting Grant (SEDMorph P.I. V.~Wild) 
and European Career Re-integration Grant (Phiz-Ev P.I. V.~Wild). 
YA acknowledges financial support from the \emph{Ram\'{o}n y Cajal} programme (RyC-2011-09461) and project 
AYA2013-47742-C4-3-P, both managed by the \emph{Ministerio de Econom\'{i}a y Competitividad}, as well as the 
`Study of Emission-Line Galaxies with Integral-Field Spectroscopy' (SELGIFS) programme, funded by the 
EU (FP7-PEOPLE-2013-IRSES-612701) within the Marie-Sklodowska-Curie Actions scheme. 
ROM acknowledges support from CAPES (Brazil) through a PDJ fellowship from project 88881.030413/2013-01, program CSF-PVE. 

\end{acknowledgements}

\bibliographystyle{aa}
\bibliography{CALIFA_DR3}

\end{document}

%% file: QCflags_def_V500.tex
\onecolumn
\begin{table}
\centering
\caption{Definition of CALIFA DR3 quality control flags for the V500 data. {\bf Numbers in square brackes provide 
the percentage
of released cubes with \textsc{OK}, \textsc{warning}, \textsc{bad}, and undefined value, respectively.}}
\label{tab:QCflags_def_V500}
\begin{tabular}{lllll}
\hline\hline\\
QC flag & QC parameters involved & \textsc{warning} condition(s) & \textsc{bad} condition(s) & Flag definition \\
\hline\hline\\
\textsc{flag\_obs\_am} & \textsc{obs\_airmass\_mean} & $>2.0$ & ... & Worst of the three parameters \\
\tiny{$[98.1\%,1.9\%,0.0\%,0.0\%]$}& \textsc{obs\_airmass\_max}  & $>2.5$ & ... & \\
                       & \textsc{obs\_airmass\_rms}  & $>0.15$ & ... & \\
\hline\\
\textsc{flag\_obs\_skymag} & \textsc{obs\_skymag\_mean} & $<19.5\,\mathrm{mag_V\,arcsec}^{-2}$ & ... & Worst of the two parameters \\
\tiny{$[95.5\%,4.5\%,0.0\%,0.0\%]$}& \textsc{obs\_skymag\_rms}  & $>0.1$ & ... & \\
\hline\\
\textsc{flag\_obs\_ext} & \textsc{obs\_ext\_mean} & $>0.30$~mag & ... & Worst of the three parameters \\
\tiny{$[65.2\%,4.0\%,0.0\%,30.8\%]$}& \textsc{obs\_ext\_max}  & $>0.35$     & ... & \\
                        & \textsc{obs\_ext\_rms}  & $>0.10$     & ... & \\
\hline
\hline\\
\textsc{flag\_red\_straylight} & \textsc{red\_meanstraylight\_max} & $>50$ counts& $>100$ & Worst of the three parameters \\
\tiny{$[84.2\%,1.7\%,3.9\%,10.2\%]$}& \textsc{red\_maxstraylight\_max} & $>75$ & $>150$ &  \\
                        & \textsc{red\_rmsstraylight\_max} & $>15$ & $>30$ &  \\
\hline\\
\textsc{flag\_red\_disp} & \textsc{red\_disp\_mean} & $>5.5$ \AA~(FWHM) & $>10 $ & Worst of the three parameters \\
\tiny{$[75.1\%,19.0\%,0.0\%,5.9\%]$}& \textsc{red\_disp\_max} & $>10.0$ & ... &  \\
                        & \textsc{red\_disp\_rms} & $>0.5$ & ... &  \\
\hline\\
\textsc{flag\_red\_cdisp} & \textsc{red\_cdisp\_mean} & $>3.0$ pixels (FWHM) & ... & Worst of the three parameters \\
\tiny{$[84.1\%,10.2\%,0.0\%,5.7\%]$}& \textsc{red\_cdisp\_max} & $\ge 4.0$ & ... &  \\
                        & \textsc{red\_cdisp\_rms} & $>0.25$ & ... &  \\
\hline\\
\textsc{flag\_red\_skylines} & \textsc{red\_res5577\_min} & $<-0.1$ counts & ... & Worst of the three parameters \\
\tiny{$[66.1\%,23.8\%,0.0\%,10.1\%]$}& \textsc{red\_res5577\_max} & $>0.1$ & ... &  \\
                             & \textsc{red\_rmsres5577\_max} & $>1.0$ & ... &  \\
\hline\\
\textsc{flag\_red\_limsb} & \textsc{red\_limsb} & $<23.0\,\mathrm{mag_{AB}\,arcsec}^{-2}$ & $<22.5$ & \\
\tiny{$[86.5\%,9.9\%,2.9\%,0.6\%]$}& & & & \\
\hline\\
\textsc{flag\_red\_errspec} & \textsc{red\_frac\_bigerr} & $>0.4$ & $>0.6$ & \\
\tiny{$[98.6\%,0.0\%,1.2\%,0.2\%]$} & & & & \\
\hline
\hline\\
\textsc{flag\_cal\_specphoto} & \textsc{cal\_qflux\_g} & $>0.06$ dex & $>0.097$ dex & Worst of the three parameters \\
\tiny{$[67.6\%,21.8\%,10.5\%,0.0\%]$}&                        & $<-0.06$ dex& $<-0.097$ dex& combined with visual checks \\
                              & \textsc{cal\_qflux\_r} & $>0.06$ dex & $>0.097$ dex & on the 30"-integrated spectrum: \\
                              &                        & $<-0.06$ dex& $<-0.097$ dex& spectral shape and comparison \\
                              & \textsc{cal\_qflux\_rms} & $>0.1$ & $>0.2$ & with SDSS photometry \\
\hline\\
\textsc{flag\_cal\_wl} & \textsc{cal\_rmsvelmean} & $>2.0~\mathrm{km~s}^{-1}$  & $>5.0$ & \\
\tiny{$[92.1\%,7.4\%,0.5\%,0.0\%]$} & & & & \\
\hline\\
\textsc{flag\_cal\_imgqual} & & & & Visual checks on\\
\tiny{$[84.8\%,14.9\%,0.0\%,0.3\%]$} &&&& synthetic broad-band image\\
\hline\\
\textsc{flag\_cal\_specqual} & & & & Visual checks on 30\arcsec-aperture\\
\tiny{$[94.6\%,5.4\%,0.0\%,0.0\%]$}&&&& integrated spectrum\\
\hline\\
\textsc{flag\_cal\_flatsdss} & & & & Visual checks on SDSSflat map,\\
\tiny{$[47.1\%,43.8\%,0.3\%,8.8\%]$}&&&& and effective SDSSflat response\\ 
&&&& from 30\arcsec-aperture integrated\\
&&&& spectrum\\
&&&& -1 if SDSSflat not applied\\
\hline\\
\textsc{flag\_cal\_registration} & & & & Visual checks on synthetic\\
\tiny{$[52.6\%,23.4\%,0.0\%,24.0\%]$}&&&& broad-band image, SDSS footprint,\\
&&&& and $\chi^2$ surface plots\\
&&&& -1 if registration relative\\
&&&& to SDSS not applied\\
\hline\hline\\

\end{tabular}
\end{table}
\twocolumn

%% file: QCflags_def_V1200.tex
\onecolumn
\begin{table}
\centering
\caption{Definition of CALIFA DR3 quality control flags for the V1200 data. {\bf Numbers in square brackes provide 
the percentage
of released cubes with \textsc{OK}, \textsc{warning}, \textsc{bad}, and undefined value, respectively.} }
\label{tab:QCflags_def_V1200}
\begin{tabular}{lllll}
\hline\hline\\
QC flag & QC parameters involved & \textsc{warning} condition(s) & \textsc{bad} condition(s) & Flag definition \\
\hline\hline\\
\textsc{flag\_obs\_am} & \textsc{obs\_airmass\_mean} & $>1.5$ & ... & Worst of the three parameters \\
\tiny{$[89.3\%,10.7\%,0.0\%,0.0\%]$}& \textsc{obs\_airmass\_max}  & $>2.0$ & ... & \\
                       & \textsc{obs\_airmass\_rms}  & $>0.15$ & ... & \\
\hline\\
\textsc{flag\_obs\_skymag} & \textsc{obs\_skymag\_mean} & $<21.5\,\mathrm{mag_V\,arcsec}^{-2}$ & ... & Worst of the two parameters \\
\tiny{$[87.6\%,12.4\%,0.0\%,0.0\%]$}& \textsc{obs\_skymag\_rms}  & $>0.1$ & ... & \\
\hline\\
\textsc{flag\_obs\_ext} & \textsc{obs\_ext\_mean} & $>0.30$~mag & ... & Worst of the three parameters \\
\tiny{$[61.4\%,1.9\%,0.0\%,36.8\%]$}& \textsc{obs\_ext\_max}  & $>0.35$     & ... & \\
                        & \textsc{obs\_ext\_rms}  & $>0.10$     & ... & \\
\hline
\hline\\
\textsc{flag\_red\_straylight} & \textsc{red\_meanstraylight\_max} & $>15$ counts& $>30$ & Worst of the three parameters \\
\tiny{$[76.9\%,13.4\%,5.6\%,4.1\%]$}& \textsc{red\_maxstraylight\_max} & $>20$ & $>50$ &  \\
                        & \textsc{red\_rmsstraylight\_max} & $>1.5$ & $>3.0$ &  \\
\hline\\
\textsc{flag\_red\_disp} & \textsc{red\_disp\_mean} & $>2.3$ \AA~(FWHM) & $>2.5$ & Worst of the three parameters \\
\tiny{$[78.1\%,19.2\%,0.0\%,2.7\%]$}& \textsc{red\_disp\_max} & $>3.0$ & ... &  \\
                        & \textsc{red\_disp\_rms} & $>0.2$ & $ ... $ &  \\
\hline\\
\textsc{flag\_red\_cdisp} & \textsc{red\_cdisp\_mean} & $>3.0$ pixels (FWHM) & ... & \\
\tiny{$[68.0\%,29.3\%,0.0\%,2.7\%]$} & & & & \\
\hline\\
\textsc{flag\_red\_skylines} & \textsc{red\_res4358\_min} & $<-0.1$ counts & ... & Worst of the three parameters \\
\tiny{$[62.6\%,33.3\%,0.0\%,4.1\%]$} & \textsc{red\_res4358\_max} & $>0.1$ & ... &  \\
                             & \textsc{red\_rmsres4358\_max} & $>0.7$ & ... &  \\
\hline\\
\textsc{flag\_red\_limsb} & \textsc{red\_limsb} & $<22.5\,\mathrm{mag_{AB}\,arcsec}^{-2}$ & $<22.0$ & \\
\tiny{$[93.8\%,4.3\%,1.9\%,0.0\%]$} & & & & \\
\hline\\
\textsc{flag\_red\_errspec} & \textsc{red\_frac\_bigerr} & $>0.4$ & $>0.6$ & \\
\tiny{$[99.8\%,0.2\%,0.0\%,0.0\%]$} & & & & \\
\hline
\hline\\
\textsc{flag\_cal\_wl} & \textsc{cal\_rmsvelmean} & $>1.0~\mathrm{km~s}^{-1}$  & $>2.0$ & \\
\tiny{$[98.1\%,1.7\%,0.0\%,0.2\%]$} & & & & \\
\hline\\
\textsc{flag\_cal\_imgqual} & & & & Visual checks on\\
\tiny{$[88.8\%,9.7\%,0.0\%,1.4\%]$}&&&& synthetic broad-band image\\
\hline\\
\textsc{flag\_cal\_specqual} & & & & Visual checks on 30\arcsec-aperture\\
\tiny{$[91.9\%,7.6\%,0.0\%,0.4\%]$}&&&& integrated spectrum\\
\hline\\
\textsc{flag\_cal\_flatsdss} & & & & Visual checks on SDSSflat map,\\
\tiny{$[55.6\%,36.0\%,0.0\%,8.5\%]$}&&&& and effective SDSSflat response\\ 
&&&& from 30\arcsec-aperture integrated\\
&&&& spectrum\\
&&&& -1 if SDSSflat not applied\\
\hline\\
\textsc{flag\_cal\_registration} & & & & Visual checks on synthetic\\
\tiny{$[33.1\%,35.5\%,0.2\%,31.2\%]$}&&&& broad-band image, SDSS footprint,\\
&&&& and $\chi^2$ surface plots\\
&&&& -1 if registration relative\\
&&&& to SDSS not applied\\
\hline\hline\\

\end{tabular}
\end{table}
\twocolumn

%% file: QCflags_def_COMB.tex
\onecolumn
\begin{table}
\centering
\caption{Definition of CALIFA DR3 quality control flags \emph{additional} for the COMB data. {\bf Numbers in square brackes provide 
the percentage
of released cubes with \textsc{OK}, \textsc{warning}, \textsc{bad}, and undefined value, respectively.} }
\label{tab:QCflags_def_COMB}
\begin{tabular}{lllll}
\hline\hline\\
QC flag & QC parameters involved & \textsc{warning} condition(s) & \textsc{bad} condition(s) & Flag definition \\
\hline\hline\\
\textsc{flag\_cal\_imgqual} & & & & Visual checks on\\
\tiny{$[76.0\%,24.0\%,0.0\%,0.0\%]$}&&&& synthetic broad-band image\\
\hline\\
\textsc{flag\_cal\_specqual} & & & & Visual checks on 30\arcsec-aperture\\
\tiny{$[96.6\%,3.4\%,0.0\%,0.0\%]$}&&&& integrated spectrum\\
\hline\\
\textsc{flag\_cal\_V1200V500} & & & & Visual checks on the match of\\
\tiny{$[79.6\%,20.4\%,0.0\%,0.0\%]$}&&&& the 30\arcsec-aperture integrated spectra\\
&&&& in V500, V1200 and resulting COMB\\
\hline\\
\textsc{flag\_cal\_wl} & \textsc{cal\_rmsvelmean} & $>2.0~\mathrm{km~s}^{-1}$  & $>5.0$ & \\
\tiny{$[97.8\%,2.2\%,0.0\%,0.0\%]$} & & & & \\
\hline\hline\\
\end{tabular}
\end{table}
\twocolumn